\documentclass[
 reprint,
 nofootinbib,
 amsmath,amssymb,
 aps,
 prl,
 lengthcheck,
]{revtex4-2}

\usepackage{graphicx}
\usepackage{dcolumn}
\usepackage{bm}
\usepackage{physics}
\usepackage{enumitem}

\newcommand{\Lab}[1]{{\ensuremath{\rm #1}}}
\newcommand{\FTIAN}{Valiev Institute of Physics and Technology, Russian Academy of Sciences,117218, Moscow, Russia}
\newcommand{\QTC}{Quantum Technology Centre, Faculty of Physics, M. V. Lomonosov Moscow State University,119991, Moscow, Russia}

\usepackage{xifthen}
\newcommand{\p}[1][]{{\ifthenelse{\isempty{#1}}{\vb*{\theta}}{\theta_{#1}}}}
\newcommand{\pe}[1][]{{\ifthenelse{\isempty{#1}}{\vb*{\tilde{\theta}}}{\tilde{\theta}_{#1}}}}

\newcommand{\HG}[1]{{\text{HG}_{#1}}}

\renewcommand{\section}[1]{\textit{#1}.---}

\begin{document}

\preprint{APS/123-QED}

\title{
Breaking Rayleigh’s curse for two unbalanced single-photon emitters: BLESS technique 
}

\author{Konstantin Katamadze}%
 \email{k.g.katamadze@gmail.com}
\affiliation{\FTIAN}%
\affiliation{\QTC}

\author{Boris Bantysh}
 \affiliation{\FTIAN}
 
\author{Andrey Chernyavskiy}
 \affiliation{\FTIAN}

\author{Yurii Bogdanov}%
\affiliation{\FTIAN}%

\author{Sergei Kulik}%
\affiliation{\QTC}

\date{\today}

\begin{abstract} 
Rayleigh's criterion states that resolving point sources below the point spread function width is impossible, with error increasing at shorter distances, known as Rayleigh’s curse. While detection mode shaping solves this for equal sources, it fails for unbalanced sources with unknown brightness ratios. We propose BLESS, a technique using Beam moduLation and Examination of Shot Statistics, breaking Rayleigh’s curse for unbalanced sources. Classical and quantum Cramér–Rao bound calculations show BLESS’s strong potential for real imaging experiments.
\end{abstract}

\maketitle 





\section{Introduction}
Fluorescence microscopy is essential for many biological applications. It uses dye molecules or quantum dots that emit photons to create images. 
The Rayleigh limit states that two emitters cannot be resolved if the distance between them is smaller than the point spread function (PSF) width $\sigma$, which is proportional to the wavelength $\lambda$~\cite{Rayleigh_1879}.

\begin{figure}[t]
\includegraphics [width=\columnwidth] {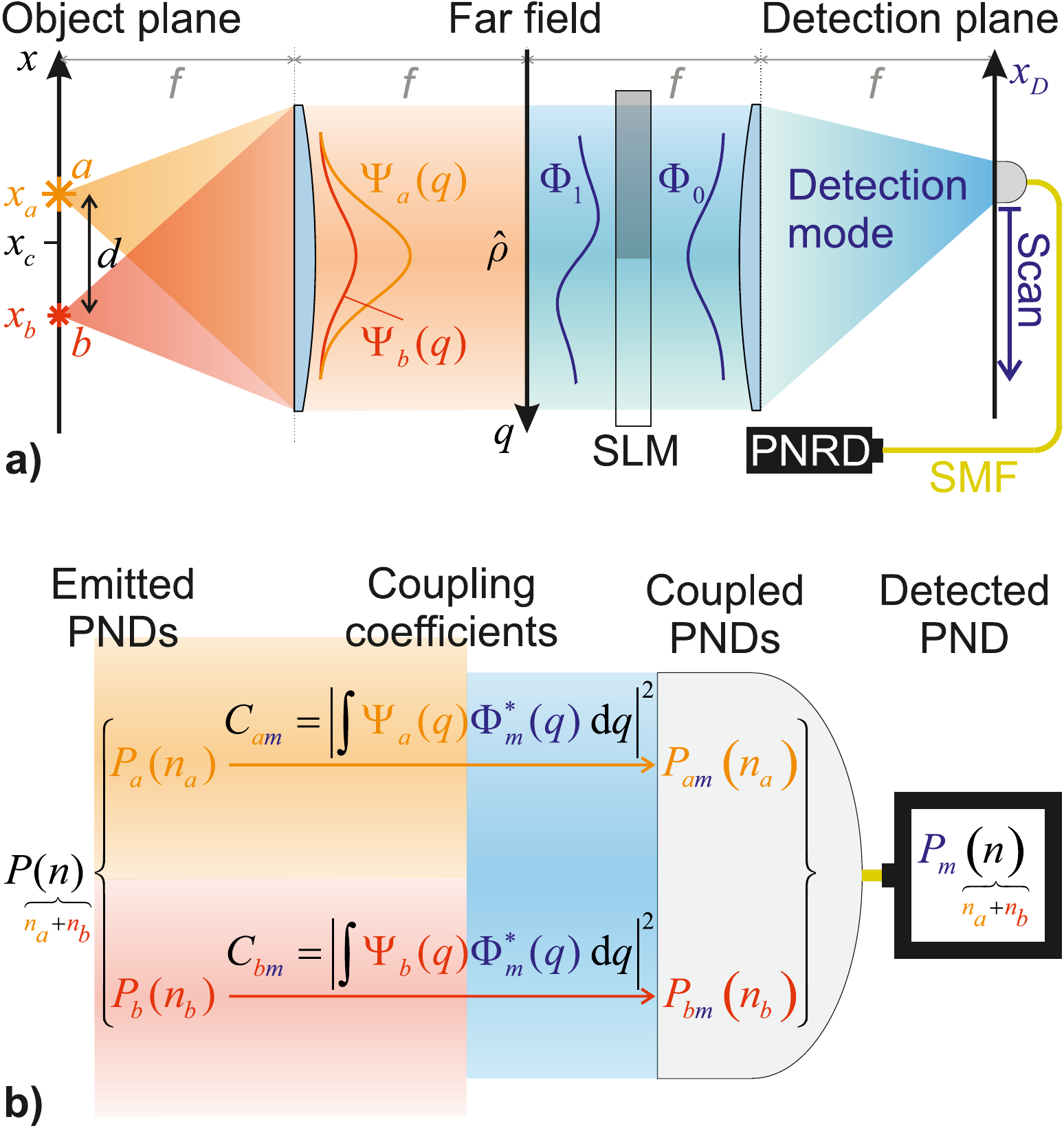}\caption{\label{fig:setup} 
(a) Principal imaging setup and (b) photon number distribution (PND) calculation scheme. 
Single-photon sources $a$ and $b$ are located at $x_a$ and $x_b$ in the object plane, with far-field spatial modes $\Psi_a$ and $\Psi_b$, and PNDs $P_a(n_a)$ and $P_b(n_b)$. The total probability of emitting $n$ photons from both sources is given by $P(n)$.
$\hat{\rho}$ represents the quantum state of the emitted light in the far field. 
SMF: single-mode fiber that forms a detection mode $\Phi_0$ centered at $x_D$ in the detection plane. 
SLM: spatial light modulator that converts $\Phi_0$ to $\Phi_1$.
$C_{sm}$ denotes the coupling coefficients between the source mode $\Psi_s$ and detection mode $\Phi_m$, which convert the emitted PNDs $P_s(n)$ to the coupled PNDs $P_{sm}(n)$ for $s = a, b$.
PNRD: photon number resolving detector, measuring the total detected PND $P_m(n)$ formed by convolving $P_{am}(n)$ and $P_{bm}(n)$.
}
\end{figure}

Several methods, such as STED microscopy~\cite{Muller_Schumann_Kraegeloh_2012}, have been developed to overcome this limit. STED reduces the PSF width through high-power donut-mode laser, depleting luminescence, but leading to dye bleaching and damage of the sample.
Another approach, used in STORM~\cite{Rust_Bates_Zhuang_2006} and PALM~\cite{Betzig_Patterson_Sougrat_Lindwasser_Olenych_Bonifacino_Davidson_Lippincott-Schwartz_Hess_2006}, involves switching dyes on and off to localize emissions, though it requires long exposures to avoid simultaneous photon emissions from nearby emitters.



Estimating the distance $d$ between nearby sources is challenging when they emit photons simultaneously, especially for $d < \sigma$. In such cases, the estimation error $\Delta_d$ increases with decreasing $d$, a phenomenon known as Rayleigh’s curse~\cite{Bettens_Van_Dyck_den_Dekker_Sijbers_van_den_Bos_1999, Van_Aert_den_Dekker_Van_Dyck_van_den_Bos_2002, Tsang_Nair_Lu_2015}. By evaluating quantum Fisher information (QFI) Tsang et al. showed the possibility to overcome Rayleigh’s curse for two \textit{equal} sources~\cite{Tsang_Nair_Lu_2015, Nair_Tsang_2016a, Nair_Tsang_2016b}. Practical protocols, like SPADE~\cite{Tsang_Nair_Lu_2015} and SLIVER~\cite{Nair_Tsang_2016a}, nearly achieve this limit but have not been applied in cases with \textit{unequal} brightness, where Rayleigh’s curse persists~\cite{Bonsma-Fisher_Tham_Ferretti_Steinberg_2019, Rehacek_Hradil_Stoklasa_Paur_Grover_Krzic_Sanchez-Soto_2017}.



Notably, these techniques ignore rare multi-photon events, when two or more photons are emitted simultaneously. However, despite their low probability, they can yield valuable information. Photon correlation measurements have enhanced resolution~\cite{Hell1995,Schwartz_Oron_2012,Schwartz_Levitt_Tenne_Itzhakov_Deutsch_Oron_2013, Gatto_Monticone_Katamadze_Traina_Moreva_Forneris_Ruo-Berchera_Olivero_Degiovanni_Brida_Genovese_2014, Tenne2019, Bartels2022, Kudyshev2023} and enabled subdiffraction localization of single-photon emitters~\cite{Israel2017,Worboys2020,Sroda2020,Li2024}, but combining this approach with detection mode shaping to overcome Rayleigh’s curse has not yet been demonstrated.


In this work, we show that including rare multi-photon events can drastically improve imaging resolution. By evaluating QFI, we for the first time demonstrate that such events can overcome Rayleigh's curse, even for unbalanced emitters.  We employ these findings in a new BLESS (Beam moduLation and Examination of Shot Statistics) technique, which combines detection mode shaping with photon number distribution (PND) measurement. This approach nearly reaches the QFI limit and is validated under realistic conditions, addressing key imaging imperfections. Compared to other super-resolution methods, BLESS is applicable to a broader range of fluorophores, causes less phototoxicity than STED, and enables faster measurements than PALM/STORM.


We begin by describing BLESS protocol in terms of measurable photon number distributions (PNDs) to calculate classical Fisher information and resolution bounds. We then develop a quantum model for PNDs and derive the ultimate precision limits using quantum Fisher information.





\section{Theoretical model}\label{sec.BLESS}
Here we analyse the imaging system, described in Fig.~\ref{fig:setup}a and derive all the PNDs, presented in Fig.~\ref{fig:setup}b in order to obtain the total detected PND $P_m(n)$, using for resolution bounds calculation.

Consider a 1D imaging problem involving two \emph{uncorrelated} and \emph{non-interfering} single-photon point sources $a$ and $b$ located in the object plane at positions $x_a$ and $x_b$, respectively.
The photon number $n$ distributions for these sources are given by:
\begin{equation}\label{eq.Ps}
    P_{s}(n_s=0) = (1-\mu_{s}),\quad P_{s}(n_s=1)=\mu_{s},
\end{equation}
where $\mu_s \in [0,1]$ is the mean photon number for each source $s=a,b$. The probability of having more than one photon from a single source, $P_s(n_s>1)$, is zero.

The total number of photons emitted from both sources follows a distribution $P(n=n_a+n_b|\mu_a,\mu_b)$:
\begin{subequations}\label{eq.P}
    \begin{align}
    P(0|\mu_a,\mu_b)=&(1-\mu_{a})(1-\mu_{b}),\\
    P(1|\mu_a,\mu_b)=&\mu_{a}(1-\mu_{b})+\mu_{b}(1-\mu_{a}),\\
    P(2|\mu_a,\mu_b)=&\mu_{a}\mu_{b}.
    \end{align}
\end{subequations}

The two-source system can be described using four key parameters:
\begin{itemize}
    \item distance $d=x_a-x_b$,
    \item mean photon number of the total PND $P(n)$ $\mu=\mu_a+\mu_b$,
    \item centroid $x_c=\left(\mu_ax_a+\mu_bx_b\right)/\mu$,
    \item relative brightness $\gamma=(\mu_a-\mu_b)/\mu\in[-1,1]$.
\end{itemize}

The light from the sources passes through a $4f$ imaging system with 1:1 magnification. Since any far-field imaging system has a limited numerical aperture (NA), this results in some loss of photons and information about the source positions.
Typically,  the aperture function can be approximated using a Gaussian function, so the light from source $s$ located at $x_s$ $(s=a,b)$ creates a Gaussian far-field electric field distribution over the transverse wave vector component $\Psi_s(q)={\rm HG}_0(q, \sigma) e^{-{\rm i} q x_s}$, where:
\begin{equation}\label{eq:HG}
{\rm HG}_m(q,\sigma)\equiv\sqrt{\frac{\sigma}{\sqrt{\pi} 2^m m!}} {\rm e}^{-\frac{{\rm i} m \pi}{2}} {\rm e}^{-\frac{\sigma^2 q^2}{2}} H_m(\sigma q)  
\end{equation}
defines the Hermite-Gaussian mode in the far-field. In the near-field, $\Psi_s(q)$ corresponds to a Gaussian PSF, $\Tilde{\Psi}_s(x) = \frac{1}{\sqrt[4]{\pi\sigma^2}} \text{e}^{-\frac{(x-x_s)^2}{2\sigma^2}}$, where $\sigma \sim \lambda/{\rm NA}$.

In the detection plane, the light is coupled into a single-mode fiber (SMF) collimator, forming a Gaussian detection mode centered at $x_D$ with waist $\sigma$:
$\Phi_{0}(q)={\rm HG}_0(q, \sigma) e^{-{\rm i} q x_D}$. By scanning the position $x_D$, the image profile can be measured.

Following the SPADE approach~\cite{Tsang_Nair_Lu_2015, Nair_Tsang_2016a, Nair_Tsang_2016b,Paur_Stoklasa_Hradil_Sanchez-Soto_Rehacek_2016, Tham_Ferretti_Steinberg_2017, Paur_Stoklasa_Grover_Krzic_Sanchez-Soto_Hradil_Rehacek_2018, Donohue2018, Boucher_Fabre_Labroille_Treps_2020}, a spatial light modulator (SLM) can be placed between the lenses to transform the Gaussian HG$_0$ detection mode into the first Hermite-Gaussian mode HG$_1$, with field distribution $\Phi_{1}(q)={\rm HG}_1(q, \sigma) e^{-{\rm i} q x_D}$. 

The probability of collecting a single photon emitted by source $s$ in the detection mode $\Phi_m$ $(m=0,1)$ is given by the coupling coefficient:
\begin{equation}\label{eq:overlap}
C_{sm}=\abs{\int \Psi_{s}^\ast(q)\Phi_m(q){\rm d}q }^2\!\!\! =\xi^m{\rm e}^{-\xi},\quad s=a,b,
\end{equation}
where $\xi\equiv\frac{(x_s-x_{D})^2}{2\sigma^2}$.

\begin{figure}[b]
\includegraphics [width=\columnwidth] {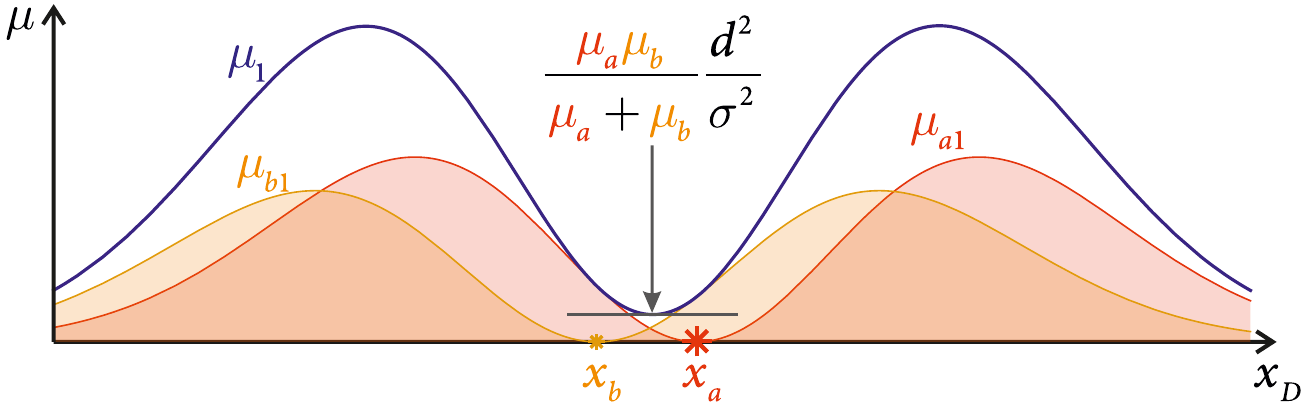}
\caption{\label{fig:minimum}
Image with the $\Phi_1$ detection mode resulting in the \Lab{HG_1} PSF. The filled orange and red curves represent the mean photon numbers $\mu_{a1}$ and $\mu_{b1}$ in the detection mode from sources $a$ and $b$, respectively. The blue curve shows the total mean number of detected photons, $\mu_1 = \mu_{a1} + \mu_{b1}$.
} 
\end{figure}

The PND of photons emitted by source $s$ and coupled into the detection mode $m$, $P_{sm}(n_s)$, follows the same form as in Eq.~\eqref{eq.Ps}, but with the mean photon number $\mu_{sm}(x_D) = \mu_s C_{sm}(x_D)$. These can be considered as images of the sources, presented in Fig.~\ref{fig:minimum}, and coupling coefficients \eqref{eq:overlap} can be considered as effective PSFs.

To measure the PND, the SMF output is connected to a photon number resolving detector (PNRD). Assuming the detector has 100\% quantum efficiency, the total detected PND $P_m(n=n_a+n_b|x_D)$ is the convolution of the coupled PNDs $P_{am}(n_a)$ and $P_{bm}(n_b)$. It has the same form as the total emitted PND~\eqref{eq.P}:
\begin{equation}\label{eq.Pd}
    P_m(n|x_D)=P(n|\mu_{am}(x_D),\mu_{bm}(x_D)).
\end{equation}

The mean detected photon number $\mu_m(x_D)$ for this distribution is:
\begin{equation}\label{eq.mu}
    \mu_m(x_D) = \mu_{am}(x_D) + \mu_{bm}(x_D),
\end{equation}
which can be considered the full image of both sources (see Fig.~\ref{fig:minimum}).


\section{Main idea}
Detection mode shaping methods~\cite{Tsang_Nair_Lu_2015, Nair_Tsang_2016a, Nair_Tsang_2016b,Paur_Stoklasa_Hradil_Sanchez-Soto_Rehacek_2016, Tham_Ferretti_Steinberg_2017, Paur_Stoklasa_Grover_Krzic_Sanchez-Soto_Hradil_Rehacek_2018, Donohue2018, Boucher_Fabre_Labroille_Treps_2020} efficiently estimate the distance $d$ between two equal sources but face challenges when dealing with sources of unknown relative brightness $\gamma$. This difficulty lies in the inability to estimate these two parameters separately. However, by independently measuring $\gamma$ through photon statistics, we can overcome this limitation and break Rayleigh's curse, even for unbalanced sources.

Indeed, for two close emitters ($d \ll \sigma$) and the \Lab{HG_1} detection mode, the mean photon number profile \eqref{eq.mu} shown in Fig.~\ref{fig:minimum} exhibits a minimum:
\begin{equation}\label{eq:minimum}
    \mu_1|_{x_D=x_c}=\frac{\mu_a \mu_b}{\mu_a+\mu_b}\frac{d^2}{\sigma^2} \equiv \frac{\mu g}{2}\frac{d^2}{\sigma^2},\quad g \equiv \frac{1-\gamma^2}{2}.
\end{equation}
With knowledge of the relative brightness $\gamma$, one can easily estimate the distance $d$ from the measured value of $\mu_1(x_c)$. From statistical error propagation theory, as detailed in Supplementary~\cite{Supplement}, the distance estimation error is given by:
\begin{equation}\label{eq:Delta.d.HG1}
    \Delta_d \approx \frac{d}{2\sqrt{\mu_1(x_c)N}} = \frac{\sigma}{\sqrt{2 \mu g N}},
\end{equation}
where $N$ denotes the number of measurements.

If no prior information about $\gamma$ is available, the problem becomes more complex. As shown in~\cite{Bonsma-Fisher_Tham_Ferretti_Steinberg_2019, Rehacek_Hradil_Stoklasa_Paur_Grover_Krzic_Sanchez-Soto_2017}, this leads to Rayleigh's curse, where $\Delta_d \propto d^{-1}$.

However, for $\mu \ll 1$, the ratio $\frac{\mu_a \mu_b}{\mu_a + \mu_b} \approx \frac{P(2)}{P(1)}$, where $P(n)$ is the total emitted PND~\eqref{eq.P}. For $d \ll \sigma$, both probabilities can be measured in the \Lab{HG_0} detection mode centered at $x_c$, where the coupling coefficients $C_{a0} \approx C_{b0} \approx 1$. The factor $g$ corresponds to the second-order correlation function $g^{(2)}(t=0)$:
\begin{equation}\label{eq:g2}
    g^{(2)} \equiv \frac{\left<n(n-1)\right>}{\left<n\right>^2} = \frac{1-\gamma^2}{2} \equiv g \approx \frac{P(2)}{[P(1)]^2}.
\end{equation}
Thus, by measuring the PND or its correlation function, one can overcome Rayleigh's curse even for unbalanced sources.

Furthermore, as derived in Supplementary~\cite{Supplement}, even measuring the PND with just the \Lab{HG_0} detection mode allows one to overcome Rayleigh's curse and estimate the distance with an error:
\begin{equation}\label{eq:Delta.d.HG0}
    \Delta_d \sim \frac{\sigma}{\mu \gamma \sqrt{g N}},
\end{equation}
for $d \ll \sigma$.

\section{Errors bounds} 
To accurately calculate the parameters' estimation errors for different measurement protocols, we apply the Cram{\'e}r--Rao bound (CRB) theory~\cite{Kendall_Stuart_1961, Cramer_1999}. For each protocol, we compute the $4\times 4$ Fisher information matrix (FIM) and then use the diagonal components of its inverse as the lower bounds for the variances of the estimated parameters: $\Delta_d^2$, $\Delta_\mu^2$, $\Delta_{x_c}^2$, and $\Delta_\gamma^2$. Detailed CRB calculations are provided in Supplementary~\cite{Supplement}.

We consider two measurement regimes. In the \textbf{PND measurement regime}, used in our BLESS protocol, the FIM is calculated based on the photon number distribution $P_m(n|d, \mu, x_c,\gamma)$ measured by the PNRD. Since the direct imaging and SPADE protocols do not account for photon statistics,  we also consider the \textbf{mean photon number measurement regime}, where we substitute the PNRD with a non-photon-number-resolving detector.This detector measures the mean photon number treated as a Gaussian random variable with an expectation value $\mu_m$ that depends on the four object parameters. We calculate the FIM based on this distribution as well.

Finally, we evaluate four different measurement protocols, presented in Fig.~\ref{fig:ESDs}~(h), each consisting of $N$ measurements:

\begin{itemize}
    \item \textbf{Direct Imaging (DI)}: Sequentially measures the mean photon number in the $\HG{0}$ mode at mesh points $x_D=\{-1, -0.8, -0.6, \dots, 0.8, 1\}\sigma$, spending an equal number of measurements across all points. This method efficiently estimates the total photon number $\mu$ and the centroid $x_c$, but it struggles to precisely estimate the relative brightness $\gamma$ and the separation $d$.
  
    \item \textbf{SPADE}: Consists of two stages. The first stage is identical to DI, while the second measures the mean photon number in the $\HG{1}$ mode at a point near $x_c$. Both stages use $N/2$ measurements. This protocol allows for the estimation of the product $gd^2$, but determining $g$ (or $\gamma$) and $d$ separately is highly challenging.
  
    \item \textbf{Examination of Shot Statistics (ESS)}: Similar to DI, but instead of just the mean photon number, it measures the full photon number distribution (PND) at each point. This approach allows for the independent estimation of all four parameters: $\mu$, $x_c$, $g$ (or $\gamma$), and $d$.
  
    \item \textbf{BLESS}: Also consists of two stages, each using $N/2$ measurements. The first stage is similar to ESS, while the second stage (like in SPADE) measures the mean photon number in the $\HG{1}$ mode near $x_c$, improving the precision of $d$ estimation. Measuring the PND in the second stage does not provide any additional benefit.
\end{itemize}

\section{Quantum limit} 
The Fisher information matrix depends on the specific measurement protocol. However, to determine the ultimate limit over all possible measurements, one needs to compute the quantum Fisher information matrix (qFIM). The qFIM is derived from the quantum state of light, which depends on the unknown parameters~\cite{Tsang_Nair_Lu_2015, Bisketzi_Branford_Datta_2019, Liu2020}.

The density operator of the field after the lens (in the far-field) produced by a single-photon source $s$ with a mean photon number $\mu_s$ is
\begin{equation}\label{eq:rhos}
    \hat \rho_s = (1 - \mu_s)\ketbra{vac} + \mu_s \ketbra{\Psi_s}, \ \  s=a,b.
\end{equation}
Here, $\ket{\Psi_s} = \int{{\rm d}q \Psi_s(q) \hat a^\dagger(q)}\ket{vac}$, and $\hat a^\dagger(q)$ creates a photon in a plane wave with transverse wave vector~$q$.

Previous works on quantum Fisher information for unbalanced emitters~\cite{Rehacek_Hradil_Stoklasa_Paur_Grover_Krzic_Sanchez-Soto_2017, Bonsma-Fisher_Tham_Ferretti_Steinberg_2019, Peng_Lu_2021} used the weak source approximation $\mu_s \ll 1$ and omitted the terms corresponding to more than one emitted photon due to their small probability. This leads to a simplified density operator for the field produced by two sources:
\begin{equation}
\begin{aligned}\label{eq.rho1}
    \hat \rho_1 &= (1 - \mu_a - \mu_b)\ketbra{vac} \\
    &+ \mu_a \ketbra{\Psi_a} + \mu_b\ketbra{\Psi_b}.
\end{aligned}
\end{equation} 
Since the state $\hat \rho_1$ contains at most one photon, the quantum Cramér--Rao bound calculated from this state is referred to as qCRB-1. However, this state does not provide information about photon statistics. While it may seem that for weak sources, the probability of detecting two photons is negligible and measuring the PND might not yield additional information, rare events can carry significant information. For example, consider measurements in PSF zeros~\cite{Paur_Stoklasa_Koutny_Rehacek_Hradil_Grover_Krzic_Sanchez-Soto_2019}, large amplification of weak values~\cite{svensson2013pedagogical}, or orthogonal measurements in quantum tomography~\cite{struchalin2018adaptive}. Therefore, accounting for weak two-photon components could be crucial.

To incorporate these rare but informative events, we define the full state as follows. Since the sources are \emph{non-interfering}, we introduce two orthogonal subspaces corresponding to each source, and write the total density operator as the following mixture of single-photon source states~\eqref{eq:rhos}:

\begin{equation}\begin{aligned}\label{eq.rho2}
    \hat\rho_2& = \frac{1}{2} \qty( \hat \rho_a \otimes \hat \rho_b + \hat \rho_b \otimes \hat \rho_a)\\&=P(0)\ketbra{vac}+P(1)\hat\rho_2^{(1)}+P(2)\hat\rho_2^{(2)},
    \end{aligned}
\end{equation}
where $\hat\rho_2^{(1)}$ and $\hat\rho_2^{(2)}$ represent the single-photon and two-photon components, respectively, and $P(n)$ is the total emitted PND, as defined in Eq.~\eqref{eq.P}.

This formulation ensures that the sources are incoherent and cannot be filtered to estimate their parameters separately. The quantum CRB derived from this state is referred to as qCRB-2. As we will show, the state $\hat \rho_2$ breaks Rayleigh’s curse even for small $\mu$, while $\hat \rho_1$ does not. Details of the derivation of the state~\eqref{eq.rho2} and the corresponding qFIM calculation are provided in Supplementary~\cite{Supplement}.

\section{Experimental imperfections} 
We address two main experimental imperfections. Firstly, non-ideal spatial modulation due to imperfections in the SLM and optical aberrations leads to cross-talk $\chi$ between the \Lab{HG_0} and \Lab{HG_1} modes. Secondly, we account for background radiation, which follows a Poissonian PND with mean photon number $\mu_\text{bg}$. Both imperfections alter the registered PND $P_m(n)$ \eqref{eq.Pd}, as detailed in Supplementary~\cite{Supplement}.

Other potential imperfections, such as optical losses and non-unitary detection efficiency, are incorporated into the parameter $\mu$, which represents the mean \textit{registered} photon number. While we assume that the sources have identical spectral and polarization properties, our technique does not depend on photon interference between the sources. Thus, if the sources can be distinguished by other degrees of freedom, localization is further simplified. Additionally, we limit our analysis to one dimension, as the problem is well-factorable for two sources.

\begin{figure*}[th]
\includegraphics [width=\textwidth] {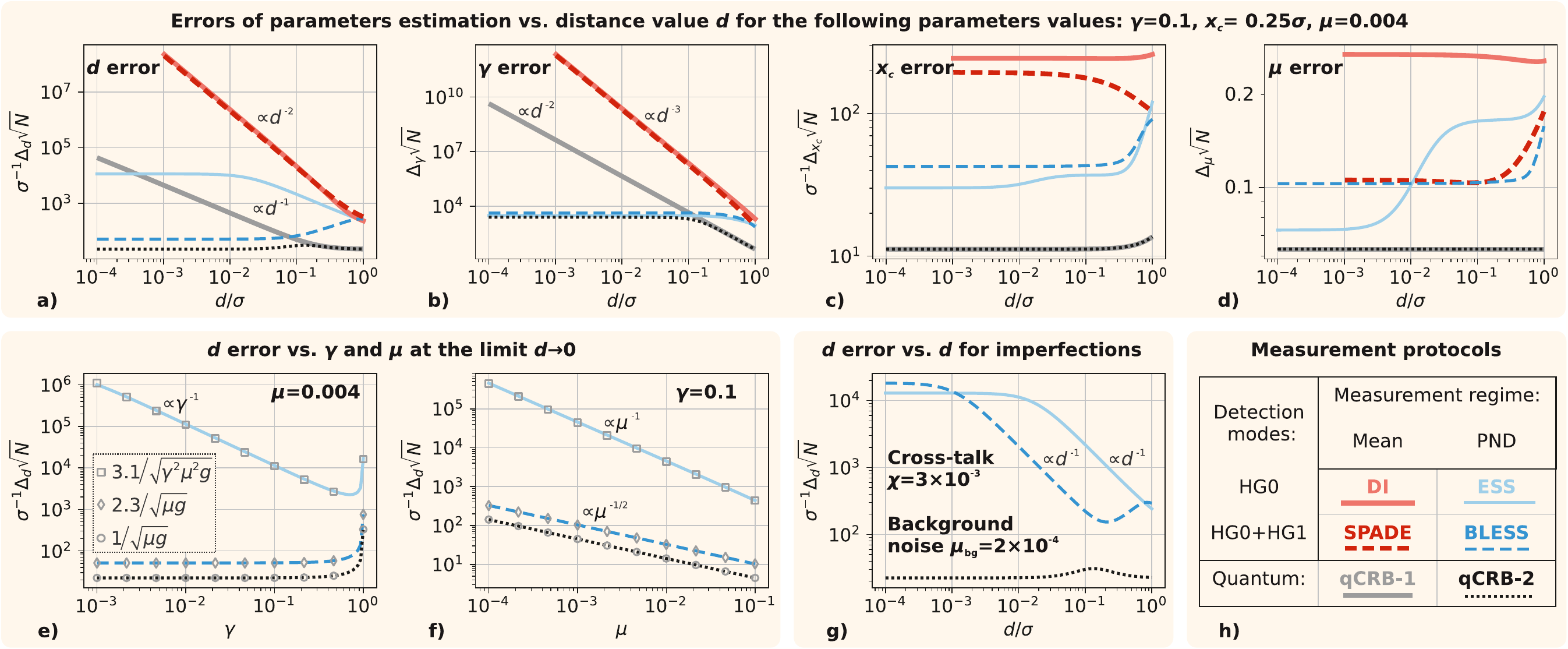}
\caption{\label{fig:ESDs} 
Normalized estimation errors $\Delta d$, $\Delta \gamma$, $\Delta x_c$, and $\Delta \mu$, calculated using the Cramér--Rao bound (CRB) for all measurement protocols (BLESS, ESS, SPADE, DI), and quantum limits for the single-photon quantum state $\hat \rho_1$ (qCRB-1) and the quantum state $\hat \rho_2$, which includes up to two photons (qCRB-2). For all plots, $\mu = 0.004$, $x_c = 0.25\sigma$, and $\gamma = 0.1$.
(a--d)~Errors for all four parameters versus $d$. The DI and SPADE lines start at $d/\sigma = 10^{-3}$, as the corresponding FIMs become nearly singular for smaller values and cannot be numerically inverted. 
\mbox{(e--f)}~Dependence of $\Delta d$ in the limit $\left(d \rightarrow 0\right)$ on $\gamma$ and $\mu$ for bounds that do not diverge. Numerical values calculated for $d = 10^{-4}\sigma$ are shown as curves. Squares, diamonds, and circles represent analytical approximations, detailed in the inset. 
\mbox{(g)}~Impact of detection mode cross-talk $(\chi = 3 \times 10^{-3})$ and background noise $(\mu_\text{bg} = 2 \times 10^{-4})$. 
(h)~Plot legends.
}
\end{figure*}

\section{Results and discussion}
Since many fluorescent microscopy techniques are first tested on stable samples, such as color centers in diamond, we evaluated our approach by simulating conditions resembling a real imaging scenario, where the aim is to resolve two nearby NV-centers. Previous studies~\cite{Gatto_Monticone_Katamadze_Traina_Moreva_Forneris_Ruo-Berchera_Olivero_Degiovanni_Brida_Genovese_2014, bogdanov2020optical} show that a single NV-center can yield a detected count rate of up to 200~kcps, corresponding to a mean photon number $\mu_s\approx 2\times 10^{-3}$ given its lifetime of 10~ns. Therefore, for two close NV-centers, we chose a total mean photon number $\mu=4\times 10^{-3}$. The other model parameters were set as $\gamma=0.1$ and $x_c=0.25\sigma$ (to significantly shift $x_c$ from the measurements' mesh points). For background radiation, we referred to~\cite{bogdanov2020optical}, where it amounted to approximately 20~kcps ($\mu_\text{bg}\approx 2\times 10^{-4}$). The cross-talk between spatial modes was set at $\chi=3\times 10^{-3}$~\cite{santamaria2023spatial}. We present the Cramér--Rao bounds (CRBs) for all measurement protocols (BLESS, ESS, SPADE, DI) and the quantum CRBs (qCRB-1 and qCRB-2) for the states $\hat \rho_1$ \eqref{eq.rho1}, consisting of no more than one photon, and $\hat \rho_2$ \eqref{eq.rho2}, consisting of up to two photons, in Fig.~\ref{fig:ESDs}.

As seen in Fig.~\ref{fig:ESDs}~(c,d), the centroid and total mean photon number estimation errors $\Delta_{x_c}$ and $\Delta_\mu$ remain small and do not increase with decreasing $d$. Some step-like features in the ESS plots are observed, but they are not significant.

Estimating $d$ and $\gamma$ is more challenging. As shown in Fig.~\ref{fig:ESDs}~(a,b), for the single-photon qCRB-1, $\Delta_d \propto d^{-1}$ and $\Delta_\gamma \propto d^{-2}$, while for direct imaging (DI), $\Delta_d \propto d^{-2}$ and $\Delta_\gamma \propto d^{-3}$, consistent with~\cite{Rehacek_Hradil_Stoklasa_Paur_Grover_Krzic_Sanchez-Soto_2017}. In contrast to the equal-source scenario, the SPADE protocol does not saturate qCRB-1 and behaves similarly to DI.

However, accounting for two-photon events qualitatively enhances the achievable resolution. The quantum CRB based on the state $\hat \rho_2$ (qCRB-2), as well as the CRBs for the ESS and BLESS protocols, which use PND measurements, are independent of $d$ for both $\Delta_d$ and $\Delta_\gamma$ estimation errors. This clearly demonstrates that this approach allows breaking Rayleigh's curse. Moreover, detection mode modulation reduces the distance estimation error $\Delta_d$ by a factor of $\sim\gamma\sqrt{\mu}$. As a result, the BLESS protocol nearly saturates the ultimate bound qCRB-2, which remains independent of $d$ except for a small peak at $d \approx 0.1$. Although we cannot fully explain this peak, it appears insignificant.

To demonstrate the protocols' efficiency for various model parameters, we present the dependence of $\Delta_d$ on $\gamma$ and $\mu$ in the limit $d\rightarrow 0$, along with their analytical approximations under the assumptions $x_c\ll \sigma$, $\mu\ll 1$ (similar to (\ref{eq:Delta.d.HG1}, \ref{eq:Delta.d.HG0}), but with a fitted factor) in Fig.~\ref{fig:ESDs}~(e,f). Analytical approximations for the error bounds of the other parameters can be found in Supplementary~\cite{Supplement}. 
An interesting and counterintuitive feature of the ESS protocol is observed: for $\gamma < 1/\sqrt{2}$, the distance estimation error scales as $\Delta_d \propto \gamma^{-1}$. This implies that more balanced sources are resolved with less accuracy, consistent with Eq.~\eqref{eq:Delta.d.HG0}, derived from statistical error propagation theory.

The impact of cross-talk $\chi$ and background noise $\mu_\text{bg}$ is shown in Fig.~\ref{fig:ESDs}~(g). We found that these imperfections may significantly decrease the efficiency of the BLESS protocol. For $d\gg \sigma \sqrt{2\frac{\mu_\text{bg}+\mu\chi}{\mu g}}$, the experimental imperfections are negligible. However, as $d$ decreases, the estimation error increases inversely with $d$ until it reaches the bound for the ESS protocol, which is robust to both imperfection types. A detailed study of the impact of imperfections is presented in Supplementary~\cite{Supplement}.

While our primary focus has been on the fundamental problem of breaking Rayleigh's curse, it is practically important to estimate the exposure time required to resolve two separated sources (achieving $\Delta_d \leq d$). Assuming the single measurement time is the same as the emitter lifetime: 10~ns, according to Fig.~\ref{fig:ESDs}~(g), the BLESS protocol with experimental imperfections can achieve a 10-fold resolution enhancement $(d=\sigma/10)$ in 10~ms, while the ESS protocol requires 1~s. This times are feasible for real imaging scenario. Under the same conditions, the DI protocol needs approximately 500~s for the same enhancement, as seen in Fig.~\ref{fig:ESDs}~(a). Note that for typical luminescence microscopy applications, the size of a single dye molecule is about 10 times smaller than the PSF width, making further resolution improvements unnecessary. A detailed analysis of practical resolution time estimation is presented in Supplementary~\cite{Supplement}.


\section{Conclusion}\label{sec.Conclusion}
In this work, we have considered the fundamental limits of fluorescence microscopy and similar imaging techniques. We have demonstrated that the common practice of neglecting multi-photon terms, frequently used for estimating optical resolution limits~\cite{Tsang_Nair_Lu_2015, Nair_Tsang_2016a, Nair_Tsang_2016b, Rehacek_Hradil_Stoklasa_Paur_Grover_Krzic_Sanchez-Soto_2017, Bonsma-Fisher_Tham_Ferretti_Steinberg_2019, Peng_Lu_2021}, can significantly reduce achievable precision. By accounting for two-photon events, we can substantially enhance resolution.

Through rigorous analysis with Cram{\'e}r--Rao bound theory (both classical and quantum), we have demonstrated for the first time the ability to break Rayleigh’s curse in a realistic scenario involving two unbalanced single-photon emitters.

We have introduced a novel emitter localization technique, BLESS, which is based on photon number distribution measurement and detection mode shaping, and which nearly achieves the fundamental resolution limits.

Our findings are directly applicable to subdiffraction localization of various single-photon sources, from trapped ions~\cite{richter_imaging_2021} and NV-centers~\cite{Gatto_Monticone_Katamadze_Traina_Moreva_Forneris_Ruo-Berchera_Olivero_Degiovanni_Brida_Genovese_2014} to dye molecules and quantum dots used in fluorescence microscopy. 

Unlike other super-resolution techniques such as STED and STORM/PALM, our approach does not use destructive high-power radiation and requires shorter exposure times, as it can resolve closely spaced sources that emit simultaneously. Moreover, it is compatible with a wider range of fluorophores, as it does not rely on luminescence depletion or blinking.



\section{Acknowledgements}
We acknowledge useful discussions with Stanislav Straupe, Egor Kovlakov, Tatiana Smirnova, Stephen Vintskevich and Ren\'e Reimann. 
This research was 
supported by the State Program no.~FFNN-2022-0016 for Valiev Institute of Physics and
Technology of RAS, and by the Russian Foundation for Basic Research (project no.~20-32-70153).

\bibliography{article}

\begin{thebibliography}{39}%
\makeatletter
\providecommand \@ifxundefined [1]{%
 \@ifx{#1\undefined}
}%
\providecommand \@ifnum [1]{%
 \ifnum #1\expandafter \@firstoftwo
 \else \expandafter \@secondoftwo
 \fi
}%
\providecommand \@ifx [1]{%
 \ifx #1\expandafter \@firstoftwo
 \else \expandafter \@secondoftwo
 \fi
}%
\providecommand \natexlab [1]{#1}%
\providecommand \enquote  [1]{``#1''}%
\providecommand \bibnamefont  [1]{#1}%
\providecommand \bibfnamefont [1]{#1}%
\providecommand \citenamefont [1]{#1}%
\providecommand \href@noop [0]{\@secondoftwo}%
\providecommand \href [0]{\begingroup \@sanitize@url \@href}%
\providecommand \@href[1]{\@@startlink{#1}\@@href}%
\providecommand \@@href[1]{\endgroup#1\@@endlink}%
\providecommand \@sanitize@url [0]{\catcode `\\12\catcode `\$12\catcode `\&12\catcode `\#12\catcode `\^12\catcode `\_12\catcode `\%12\relax}%
\providecommand \@@startlink[1]{}%
\providecommand \@@endlink[0]{}%
\providecommand \url  [0]{\begingroup\@sanitize@url \@url }%
\providecommand \@url [1]{\endgroup\@href {#1}{\urlprefix }}%
\providecommand \urlprefix  [0]{URL }%
\providecommand \Eprint [0]{\href }%
\providecommand \doibase [0]{https://doi.org/}%
\providecommand \selectlanguage [0]{\@gobble}%
\providecommand \bibinfo  [0]{\@secondoftwo}%
\providecommand \bibfield  [0]{\@secondoftwo}%
\providecommand \translation [1]{[#1]}%
\providecommand \BibitemOpen [0]{}%
\providecommand \bibitemStop [0]{}%
\providecommand \bibitemNoStop [0]{.\EOS\space}%
\providecommand \EOS [0]{\spacefactor3000\relax}%
\providecommand \BibitemShut  [1]{\csname bibitem#1\endcsname}%
\let\auto@bib@innerbib\@empty
\bibitem [{\citenamefont {Rayleigh}(1879)}]{Rayleigh_1879}%
  \BibitemOpen
  \bibfield  {author} {\bibinfo {author} {\bibfnamefont {F.~R.~S.}\ \bibnamefont {Rayleigh}},\ }\bibfield  {title} {\bibinfo {title} {Xxxi. investigations in optics, with special reference to the spectroscope},\ }\href {https://doi.org/10.1080/14786447908639684} {\bibfield  {journal} {\bibinfo  {journal} {The London, Edinburgh, and Dublin Philosophical Magazine and Journal of Science}\ }\textbf {\bibinfo {volume} {8}},\ \bibinfo {pages} {261–274} (\bibinfo {year} {1879})}\BibitemShut {NoStop}%
\bibitem [{\citenamefont {Müller}\ \emph {et~al.}(2012)\citenamefont {Müller}, \citenamefont {Schumann},\ and\ \citenamefont {Kraegeloh}}]{Muller_Schumann_Kraegeloh_2012}%
  \BibitemOpen
  \bibfield  {author} {\bibinfo {author} {\bibfnamefont {T.}~\bibnamefont {Müller}}, \bibinfo {author} {\bibfnamefont {C.}~\bibnamefont {Schumann}},\ and\ \bibinfo {author} {\bibfnamefont {A.}~\bibnamefont {Kraegeloh}},\ }\bibfield  {title} {\bibinfo {title} {Sted microscopy and its applications: New insights into cellular processes on the nanoscale},\ }\href {https://doi.org/10.1002/cphc.201100986} {\bibfield  {journal} {\bibinfo  {journal} {ChemPhysChem}\ }\textbf {\bibinfo {volume} {13}},\ \bibinfo {pages} {1986–2000} (\bibinfo {year} {2012})}\BibitemShut {NoStop}%
\bibitem [{\citenamefont {Rust}\ \emph {et~al.}(2006)\citenamefont {Rust}, \citenamefont {Bates},\ and\ \citenamefont {Zhuang}}]{Rust_Bates_Zhuang_2006}%
  \BibitemOpen
  \bibfield  {author} {\bibinfo {author} {\bibfnamefont {M.~J.}\ \bibnamefont {Rust}}, \bibinfo {author} {\bibfnamefont {M.}~\bibnamefont {Bates}},\ and\ \bibinfo {author} {\bibfnamefont {X.}~\bibnamefont {Zhuang}},\ }\bibfield  {title} {\bibinfo {title} {Sub-diffraction-limit imaging by stochastic optical reconstruction microscopy (storm)},\ }\href {https://doi.org/10.1038/nmeth929} {\bibfield  {journal} {\bibinfo  {journal} {Nature Methods}\ }\textbf {\bibinfo {volume} {3}},\ \bibinfo {pages} {793–796} (\bibinfo {year} {2006})}\BibitemShut {NoStop}%
\bibitem [{\citenamefont {Betzig}\ \emph {et~al.}(2006)\citenamefont {Betzig}, \citenamefont {Patterson}, \citenamefont {Sougrat}, \citenamefont {Lindwasser}, \citenamefont {Olenych}, \citenamefont {Bonifacino}, \citenamefont {Davidson}, \citenamefont {Lippincott-Schwartz},\ and\ \citenamefont {Hess}}]{Betzig_Patterson_Sougrat_Lindwasser_Olenych_Bonifacino_Davidson_Lippincott-Schwartz_Hess_2006}%
  \BibitemOpen
  \bibfield  {author} {\bibinfo {author} {\bibfnamefont {E.}~\bibnamefont {Betzig}}, \bibinfo {author} {\bibfnamefont {G.~H.}\ \bibnamefont {Patterson}}, \bibinfo {author} {\bibfnamefont {R.}~\bibnamefont {Sougrat}}, \bibinfo {author} {\bibfnamefont {O.~W.}\ \bibnamefont {Lindwasser}}, \bibinfo {author} {\bibfnamefont {S.}~\bibnamefont {Olenych}}, \bibinfo {author} {\bibfnamefont {J.~S.}\ \bibnamefont {Bonifacino}}, \bibinfo {author} {\bibfnamefont {M.~W.}\ \bibnamefont {Davidson}}, \bibinfo {author} {\bibfnamefont {J.}~\bibnamefont {Lippincott-Schwartz}},\ and\ \bibinfo {author} {\bibfnamefont {H.~F.}\ \bibnamefont {Hess}},\ }\bibfield  {title} {\bibinfo {title} {Imaging intracellular fluorescent proteins at nanometer resolution},\ }\href {https://doi.org/10.1126/science.1127344} {\bibfield  {journal} {\bibinfo  {journal} {Science}\ }\textbf {\bibinfo {volume} {313}},\ \bibinfo {pages} {1642–1645} (\bibinfo {year} {2006})}\BibitemShut {NoStop}%
\bibitem [{\citenamefont {Bettens}\ \emph {et~al.}(1999)\citenamefont {Bettens}, \citenamefont {Van~Dyck}, \citenamefont {den Dekker}, \citenamefont {Sijbers},\ and\ \citenamefont {van~den Bos}}]{Bettens_Van_Dyck_den_Dekker_Sijbers_van_den_Bos_1999}%
  \BibitemOpen
  \bibfield  {author} {\bibinfo {author} {\bibfnamefont {E.}~\bibnamefont {Bettens}}, \bibinfo {author} {\bibfnamefont {D.}~\bibnamefont {Van~Dyck}}, \bibinfo {author} {\bibfnamefont {A.}~\bibnamefont {den Dekker}}, \bibinfo {author} {\bibfnamefont {J.}~\bibnamefont {Sijbers}},\ and\ \bibinfo {author} {\bibfnamefont {A.}~\bibnamefont {van~den Bos}},\ }\bibfield  {title} {\bibinfo {title} {Model-based two-object resolution from observations having counting statistics},\ }\href {https://doi.org/10.1016/S0304-3991(99)00006-6} {\bibfield  {journal} {\bibinfo  {journal} {Ultramicroscopy}\ }\textbf {\bibinfo {volume} {77}},\ \bibinfo {pages} {37–48} (\bibinfo {year} {1999})}\BibitemShut {NoStop}%
\bibitem [{\citenamefont {Van~Aert}\ \emph {et~al.}(2002)\citenamefont {Van~Aert}, \citenamefont {den Dekker}, \citenamefont {Van~Dyck},\ and\ \citenamefont {van~den Bos}}]{Van_Aert_den_Dekker_Van_Dyck_van_den_Bos_2002}%
  \BibitemOpen
  \bibfield  {author} {\bibinfo {author} {\bibfnamefont {S.}~\bibnamefont {Van~Aert}}, \bibinfo {author} {\bibfnamefont {A.}~\bibnamefont {den Dekker}}, \bibinfo {author} {\bibfnamefont {D.}~\bibnamefont {Van~Dyck}},\ and\ \bibinfo {author} {\bibfnamefont {A.}~\bibnamefont {van~den Bos}},\ }\bibfield  {title} {\bibinfo {title} {High-resolution electron microscopy and electron tomography: resolution versus precision},\ }\href {https://doi.org/10.1016/S1047-8477(02)00016-3} {\bibfield  {journal} {\bibinfo  {journal} {Journal of Structural Biology}\ }\textbf {\bibinfo {volume} {138}},\ \bibinfo {pages} {21–33} (\bibinfo {year} {2002})}\BibitemShut {NoStop}%
\bibitem [{\citenamefont {Tsang}\ \emph {et~al.}(2016)\citenamefont {Tsang}, \citenamefont {Nair},\ and\ \citenamefont {Lu}}]{Tsang_Nair_Lu_2015}%
  \BibitemOpen
  \bibfield  {author} {\bibinfo {author} {\bibfnamefont {M.}~\bibnamefont {Tsang}}, \bibinfo {author} {\bibfnamefont {R.}~\bibnamefont {Nair}},\ and\ \bibinfo {author} {\bibfnamefont {X.-M.}\ \bibnamefont {Lu}},\ }\bibfield  {title} {\bibinfo {title} {Quantum theory of superresolution for two incoherent optical point sources},\ }\href {https://doi.org/10.1103/PhysRevX.6.031033} {\bibfield  {journal} {\bibinfo  {journal} {Physical Review X}\ }\textbf {\bibinfo {volume} {6}},\ \bibinfo {pages} {031033} (\bibinfo {year} {2016})}\BibitemShut {NoStop}%
\bibitem [{\citenamefont {Nair}\ and\ \citenamefont {Tsang}(2016{\natexlab{a}})}]{Nair_Tsang_2016a}%
  \BibitemOpen
  \bibfield  {author} {\bibinfo {author} {\bibfnamefont {R.}~\bibnamefont {Nair}}\ and\ \bibinfo {author} {\bibfnamefont {M.}~\bibnamefont {Tsang}},\ }\bibfield  {title} {\bibinfo {title} {Interferometric superlocalization of two incoherent optical point sources},\ }\href {https://doi.org/10.1364/oe.24.003684} {\bibfield  {journal} {\bibinfo  {journal} {Optics Express}\ }\textbf {\bibinfo {volume} {24}},\ \bibinfo {pages} {3684} (\bibinfo {year} {2016}{\natexlab{a}})}\BibitemShut {NoStop}%
\bibitem [{\citenamefont {Nair}\ and\ \citenamefont {Tsang}(2016{\natexlab{b}})}]{Nair_Tsang_2016b}%
  \BibitemOpen
  \bibfield  {author} {\bibinfo {author} {\bibfnamefont {R.}~\bibnamefont {Nair}}\ and\ \bibinfo {author} {\bibfnamefont {M.}~\bibnamefont {Tsang}},\ }\bibfield  {title} {\bibinfo {title} {Far-field superresolution of thermal electromagnetic sources at the quantum limit},\ }\href {https://doi.org/10.1103/PhysRevLett.117.190801} {\bibfield  {journal} {\bibinfo  {journal} {Physical Review Letters}\ }\textbf {\bibinfo {volume} {117}},\ \bibinfo {pages} {190801} (\bibinfo {year} {2016}{\natexlab{b}})}\BibitemShut {NoStop}%
\bibitem [{\citenamefont {Bonsma-Fisher}\ \emph {et~al.}(2019)\citenamefont {Bonsma-Fisher}, \citenamefont {Tham}, \citenamefont {Ferretti},\ and\ \citenamefont {Steinberg}}]{Bonsma-Fisher_Tham_Ferretti_Steinberg_2019}%
  \BibitemOpen
  \bibfield  {author} {\bibinfo {author} {\bibfnamefont {K.~A.~G.}\ \bibnamefont {Bonsma-Fisher}}, \bibinfo {author} {\bibfnamefont {W.-K.}\ \bibnamefont {Tham}}, \bibinfo {author} {\bibfnamefont {H.}~\bibnamefont {Ferretti}},\ and\ \bibinfo {author} {\bibfnamefont {A.~M.}\ \bibnamefont {Steinberg}},\ }\bibfield  {title} {\bibinfo {title} {Realistic sub-rayleigh imaging with phase-sensitive measurements},\ }\href {https://doi.org/10.1088/1367-2630/ab3d97} {\bibfield  {journal} {\bibinfo  {journal} {New Journal of Physics}\ }\textbf {\bibinfo {volume} {21}},\ \bibinfo {pages} {093010} (\bibinfo {year} {2019})}\BibitemShut {NoStop}%
\bibitem [{\citenamefont {Řehaček}\ \emph {et~al.}(2017)\citenamefont {Řehaček}, \citenamefont {Hradil}, \citenamefont {Stoklasa}, \citenamefont {Paúr}, \citenamefont {Grover}, \citenamefont {Krzic},\ and\ \citenamefont {Sánchez-Soto}}]{Rehacek_Hradil_Stoklasa_Paur_Grover_Krzic_Sanchez-Soto_2017}%
  \BibitemOpen
  \bibfield  {author} {\bibinfo {author} {\bibfnamefont {J.}~\bibnamefont {Řehaček}}, \bibinfo {author} {\bibfnamefont {Z.}~\bibnamefont {Hradil}}, \bibinfo {author} {\bibfnamefont {B.}~\bibnamefont {Stoklasa}}, \bibinfo {author} {\bibfnamefont {M.}~\bibnamefont {Paúr}}, \bibinfo {author} {\bibfnamefont {J.}~\bibnamefont {Grover}}, \bibinfo {author} {\bibfnamefont {A.}~\bibnamefont {Krzic}},\ and\ \bibinfo {author} {\bibfnamefont {L.~L.}\ \bibnamefont {Sánchez-Soto}},\ }\bibfield  {title} {\bibinfo {title} {Multiparameter quantum metrology of incoherent point sources: Towards realistic superresolution},\ }\href {https://doi.org/10.1103/PhysRevA.96.062107} {\bibfield  {journal} {\bibinfo  {journal} {Physical Review A}\ }\textbf {\bibinfo {volume} {96}},\ \bibinfo {pages} {062107} (\bibinfo {year} {2017})}\BibitemShut {NoStop}%
\bibitem [{\citenamefont {Hell}\ \emph {et~al.}(1995)\citenamefont {Hell}, \citenamefont {Soukka},\ and\ \citenamefont {H{\"{a}}nninen}}]{Hell1995}%
  \BibitemOpen
  \bibfield  {author} {\bibinfo {author} {\bibfnamefont {S.~W.}\ \bibnamefont {Hell}}, \bibinfo {author} {\bibfnamefont {J.}~\bibnamefont {Soukka}},\ and\ \bibinfo {author} {\bibfnamefont {P.~E.}\ \bibnamefont {H{\"{a}}nninen}},\ }\bibfield  {title} {\bibinfo {title} {{Two‐ and multiphoton detection as an imaging mode and means of increasing the resolution in far‐field light microscopy: A study based on photon‐optics}},\ }\href {https://doi.org/10.1002/1361-6374(199506)3:2<64::AID-BIO2>3.3.CO;2-F} {\bibfield  {journal} {\bibinfo  {journal} {Bioimaging}\ }\textbf {\bibinfo {volume} {3}},\ \bibinfo {pages} {64} (\bibinfo {year} {1995})}\BibitemShut {NoStop}%
\bibitem [{\citenamefont {Schwartz}\ and\ \citenamefont {Oron}(2012)}]{Schwartz_Oron_2012}%
  \BibitemOpen
  \bibfield  {author} {\bibinfo {author} {\bibfnamefont {O.}~\bibnamefont {Schwartz}}\ and\ \bibinfo {author} {\bibfnamefont {D.}~\bibnamefont {Oron}},\ }\bibfield  {title} {\bibinfo {title} {Improved resolution in fluorescence microscopy using quantum correlations},\ }\href {https://doi.org/10.1103/PhysRevA.85.033812} {\bibfield  {journal} {\bibinfo  {journal} {Phys. Rev. A}\ }\textbf {\bibinfo {volume} {85}},\ \bibinfo {pages} {33812} (\bibinfo {year} {2012})}\BibitemShut {NoStop}%
\bibitem [{\citenamefont {Schwartz}\ \emph {et~al.}(2013)\citenamefont {Schwartz}, \citenamefont {Levitt}, \citenamefont {Tenne}, \citenamefont {Itzhakov}, \citenamefont {Deutsch},\ and\ \citenamefont {Oron}}]{Schwartz_Levitt_Tenne_Itzhakov_Deutsch_Oron_2013}%
  \BibitemOpen
  \bibfield  {author} {\bibinfo {author} {\bibfnamefont {O.}~\bibnamefont {Schwartz}}, \bibinfo {author} {\bibfnamefont {J.~M.}\ \bibnamefont {Levitt}}, \bibinfo {author} {\bibfnamefont {R.}~\bibnamefont {Tenne}}, \bibinfo {author} {\bibfnamefont {S.}~\bibnamefont {Itzhakov}}, \bibinfo {author} {\bibfnamefont {Z.}~\bibnamefont {Deutsch}},\ and\ \bibinfo {author} {\bibfnamefont {D.}~\bibnamefont {Oron}},\ }\bibfield  {title} {\bibinfo {title} {Superresolution microscopy with quantum emitters.},\ }\href {https://doi.org/10.1021/nl402552m} {\bibfield  {journal} {\bibinfo  {journal} {Nano Lett.}\ }\textbf {\bibinfo {volume} {13}},\ \bibinfo {pages} {5832–5836} (\bibinfo {year} {2013})}\BibitemShut {NoStop}%
\bibitem [{\citenamefont {Gatto~Monticone}\ \emph {et~al.}(2014)\citenamefont {Gatto~Monticone}, \citenamefont {Katamadze}, \citenamefont {Traina}, \citenamefont {Moreva}, \citenamefont {Forneris}, \citenamefont {Ruo-Berchera}, \citenamefont {Olivero}, \citenamefont {Degiovanni}, \citenamefont {Brida},\ and\ \citenamefont {Genovese}}]{Gatto_Monticone_Katamadze_Traina_Moreva_Forneris_Ruo-Berchera_Olivero_Degiovanni_Brida_Genovese_2014}%
  \BibitemOpen
  \bibfield  {author} {\bibinfo {author} {\bibfnamefont {D.}~\bibnamefont {Gatto~Monticone}}, \bibinfo {author} {\bibfnamefont {K.}~\bibnamefont {Katamadze}}, \bibinfo {author} {\bibfnamefont {P.}~\bibnamefont {Traina}}, \bibinfo {author} {\bibfnamefont {E.}~\bibnamefont {Moreva}}, \bibinfo {author} {\bibfnamefont {J.}~\bibnamefont {Forneris}}, \bibinfo {author} {\bibfnamefont {I.}~\bibnamefont {Ruo-Berchera}}, \bibinfo {author} {\bibfnamefont {P.}~\bibnamefont {Olivero}}, \bibinfo {author} {\bibfnamefont {I.~P.}\ \bibnamefont {Degiovanni}}, \bibinfo {author} {\bibfnamefont {G.}~\bibnamefont {Brida}},\ and\ \bibinfo {author} {\bibfnamefont {M.}~\bibnamefont {Genovese}},\ }\bibfield  {title} {\bibinfo {title} {Beating the abbe diffraction limit in confocal microscopy via nonclassical photon statistics},\ }\href {https://doi.org/10.1103/PhysRevLett.113.143602} {\bibfield  {journal} {\bibinfo  {journal} {Physical Review Letters}\ }\textbf {\bibinfo {volume} {113}},\ \bibinfo {pages} {143602} (\bibinfo {year}
  {2014})}\BibitemShut {NoStop}%
\bibitem [{\citenamefont {Tenne}\ \emph {et~al.}(2019)\citenamefont {Tenne}, \citenamefont {Rossman}, \citenamefont {Rephael}, \citenamefont {Israel}, \citenamefont {Krupinski-Ptaszek}, \citenamefont {Lapkiewicz}, \citenamefont {Silberberg},\ and\ \citenamefont {Oron}}]{Tenne2019}%
  \BibitemOpen
  \bibfield  {author} {\bibinfo {author} {\bibfnamefont {R.}~\bibnamefont {Tenne}}, \bibinfo {author} {\bibfnamefont {U.}~\bibnamefont {Rossman}}, \bibinfo {author} {\bibfnamefont {B.}~\bibnamefont {Rephael}}, \bibinfo {author} {\bibfnamefont {Y.}~\bibnamefont {Israel}}, \bibinfo {author} {\bibfnamefont {A.}~\bibnamefont {Krupinski-Ptaszek}}, \bibinfo {author} {\bibfnamefont {R.}~\bibnamefont {Lapkiewicz}}, \bibinfo {author} {\bibfnamefont {Y.}~\bibnamefont {Silberberg}},\ and\ \bibinfo {author} {\bibfnamefont {D.}~\bibnamefont {Oron}},\ }\bibfield  {title} {\bibinfo {title} {{Super-resolution enhancement by quantum image scanning microscopy}},\ }\href {https://doi.org/10.1038/s41566-018-0324-z} {\bibfield  {journal} {\bibinfo  {journal} {Nature Photonics}\ }\textbf {\bibinfo {volume} {13}},\ \bibinfo {pages} {116} (\bibinfo {year} {2019})},\ \Eprint {https://arxiv.org/abs/1806.07661} {arXiv:1806.07661} \BibitemShut {NoStop}%
\bibitem [{\citenamefont {Bartels}\ \emph {et~al.}(2022)\citenamefont {Bartels}, \citenamefont {Murray}, \citenamefont {Field},\ and\ \citenamefont {Squier}}]{Bartels2022}%
  \BibitemOpen
  \bibfield  {author} {\bibinfo {author} {\bibfnamefont {R.~A.}\ \bibnamefont {Bartels}}, \bibinfo {author} {\bibfnamefont {G.}~\bibnamefont {Murray}}, \bibinfo {author} {\bibfnamefont {J.}~\bibnamefont {Field}},\ and\ \bibinfo {author} {\bibfnamefont {J.}~\bibnamefont {Squier}},\ }\bibfield  {title} {\bibinfo {title} {{Super-Resolution Imaging by Computationally Fusing Quantum and Classical Optical Information}},\ }\href {https://doi.org/10.34133/icomputing.0003} {\bibfield  {journal} {\bibinfo  {journal} {Intelligent Computing}\ }\textbf {\bibinfo {volume} {2022}},\ \bibinfo {pages} {1} (\bibinfo {year} {2022})}\BibitemShut {NoStop}%
\bibitem [{\citenamefont {Kudyshev}\ \emph {et~al.}(2023)\citenamefont {Kudyshev}, \citenamefont {Sychev}, \citenamefont {Martin}, \citenamefont {Yesilyurt}, \citenamefont {Bogdanov}, \citenamefont {Xu}, \citenamefont {Chen}, \citenamefont {Kildishev}, \citenamefont {Boltasseva},\ and\ \citenamefont {Shalaev}}]{Kudyshev2023}%
  \BibitemOpen
  \bibfield  {author} {\bibinfo {author} {\bibfnamefont {Z.~A.}\ \bibnamefont {Kudyshev}}, \bibinfo {author} {\bibfnamefont {D.}~\bibnamefont {Sychev}}, \bibinfo {author} {\bibfnamefont {Z.}~\bibnamefont {Martin}}, \bibinfo {author} {\bibfnamefont {O.}~\bibnamefont {Yesilyurt}}, \bibinfo {author} {\bibfnamefont {S.~I.}\ \bibnamefont {Bogdanov}}, \bibinfo {author} {\bibfnamefont {X.}~\bibnamefont {Xu}}, \bibinfo {author} {\bibfnamefont {P.-G.}\ \bibnamefont {Chen}}, \bibinfo {author} {\bibfnamefont {A.~V.}\ \bibnamefont {Kildishev}}, \bibinfo {author} {\bibfnamefont {A.}~\bibnamefont {Boltasseva}},\ and\ \bibinfo {author} {\bibfnamefont {V.~M.}\ \bibnamefont {Shalaev}},\ }\bibfield  {title} {\bibinfo {title} {{Machine learning assisted quantum super-resolution microscopy}},\ }\href {https://doi.org/10.1038/s41467-023-40506-4} {\bibfield  {journal} {\bibinfo  {journal} {Nature Communications}\ }\textbf {\bibinfo {volume} {14}},\ \bibinfo {pages} {4828} (\bibinfo {year} {2023})}\BibitemShut {NoStop}%
\bibitem [{\citenamefont {Israel}\ \emph {et~al.}(2017)\citenamefont {Israel}, \citenamefont {Tenne}, \citenamefont {Oron},\ and\ \citenamefont {Silberberg}}]{Israel2017}%
  \BibitemOpen
  \bibfield  {author} {\bibinfo {author} {\bibfnamefont {Y.}~\bibnamefont {Israel}}, \bibinfo {author} {\bibfnamefont {R.}~\bibnamefont {Tenne}}, \bibinfo {author} {\bibfnamefont {D.}~\bibnamefont {Oron}},\ and\ \bibinfo {author} {\bibfnamefont {Y.}~\bibnamefont {Silberberg}},\ }\bibfield  {title} {\bibinfo {title} {{Quantum correlation enhanced super-resolution localization microscopy enabled by a fibre bundle camera}},\ }\href {https://doi.org/10.1038/ncomms14786} {\bibfield  {journal} {\bibinfo  {journal} {Nature Communications}\ }\textbf {\bibinfo {volume} {8}},\ \bibinfo {pages} {14786} (\bibinfo {year} {2017})},\ \Eprint {https://arxiv.org/abs/1609.00312} {arXiv:1609.00312} \BibitemShut {NoStop}%
\bibitem [{\citenamefont {Worboys}\ \emph {et~al.}(2020)\citenamefont {Worboys}, \citenamefont {Drumm},\ and\ \citenamefont {Greentree}}]{Worboys2020}%
  \BibitemOpen
  \bibfield  {author} {\bibinfo {author} {\bibfnamefont {J.~G.}\ \bibnamefont {Worboys}}, \bibinfo {author} {\bibfnamefont {D.~W.}\ \bibnamefont {Drumm}},\ and\ \bibinfo {author} {\bibfnamefont {A.~D.}\ \bibnamefont {Greentree}},\ }\bibfield  {title} {\bibinfo {title} {{Quantum multilateration: Subdiffraction emitter pair localization via three spatially separate Hanbury Brown and Twiss measurements}},\ }\href {https://doi.org/10.1103/PhysRevA.101.013810} {\bibfield  {journal} {\bibinfo  {journal} {Physical Review A}\ }\textbf {\bibinfo {volume} {101}},\ \bibinfo {pages} {13810} (\bibinfo {year} {2020})}\BibitemShut {NoStop}%
\bibitem [{\citenamefont {Sroda}\ \emph {et~al.}(2020)\citenamefont {Sroda}, \citenamefont {Makowski}, \citenamefont {Tenne}, \citenamefont {Rossman}, \citenamefont {Lubin}, \citenamefont {Oron},\ and\ \citenamefont {Lapkiewicz}}]{Sroda2020}%
  \BibitemOpen
  \bibfield  {author} {\bibinfo {author} {\bibfnamefont {A.}~\bibnamefont {Sroda}}, \bibinfo {author} {\bibfnamefont {A.}~\bibnamefont {Makowski}}, \bibinfo {author} {\bibfnamefont {R.}~\bibnamefont {Tenne}}, \bibinfo {author} {\bibfnamefont {U.}~\bibnamefont {Rossman}}, \bibinfo {author} {\bibfnamefont {G.}~\bibnamefont {Lubin}}, \bibinfo {author} {\bibfnamefont {D.}~\bibnamefont {Oron}},\ and\ \bibinfo {author} {\bibfnamefont {R.}~\bibnamefont {Lapkiewicz}},\ }\bibfield  {title} {\bibinfo {title} {{SOFISM: Super-resolution optical fluctuation image scanning microscopy}},\ }\href {https://doi.org/10.1364/optica.399600} {\bibfield  {journal} {\bibinfo  {journal} {Optica}\ }\textbf {\bibinfo {volume} {7}},\ \bibinfo {pages} {1308} (\bibinfo {year} {2020})},\ \Eprint {https://arxiv.org/abs/2002.00182} {arXiv:2002.00182} \BibitemShut {NoStop}%
\bibitem [{\citenamefont {Li}\ \emph {et~al.}(2024)\citenamefont {Li}, \citenamefont {Li}, \citenamefont {Sun}, \citenamefont {Moran}, \citenamefont {Brown}, \citenamefont {Gibson},\ and\ \citenamefont {Greentree}}]{Li2024}%
  \BibitemOpen
  \bibfield  {author} {\bibinfo {author} {\bibfnamefont {S.}~\bibnamefont {Li}}, \bibinfo {author} {\bibfnamefont {W.}~\bibnamefont {Li}}, \bibinfo {author} {\bibfnamefont {Q.}~\bibnamefont {Sun}}, \bibinfo {author} {\bibfnamefont {B.}~\bibnamefont {Moran}}, \bibinfo {author} {\bibfnamefont {T.~C.}\ \bibnamefont {Brown}}, \bibinfo {author} {\bibfnamefont {B.~C.}\ \bibnamefont {Gibson}},\ and\ \bibinfo {author} {\bibfnamefont {A.~D.}\ \bibnamefont {Greentree}},\ }\bibfield  {title} {\bibinfo {title} {{Localising two sub-diffraction emitters in 3D using quantum correlation microscopy}},\ }\href {https://doi.org/10.1088/1367-2630/ad31d4} {\bibfield  {journal} {\bibinfo  {journal} {New Journal of Physics}\ }\textbf {\bibinfo {volume} {26}},\ \bibinfo {pages} {033036} (\bibinfo {year} {2024})},\ \Eprint {https://arxiv.org/abs/2310.02585} {arXiv:2310.02585} \BibitemShut {NoStop}%
\bibitem [{\citenamefont {Paúr}\ \emph {et~al.}(2016)\citenamefont {Paúr}, \citenamefont {Stoklasa}, \citenamefont {Hradil}, \citenamefont {Sánchez-Soto},\ and\ \citenamefont {Rehacek}}]{Paur_Stoklasa_Hradil_Sanchez-Soto_Rehacek_2016}%
  \BibitemOpen
  \bibfield  {author} {\bibinfo {author} {\bibfnamefont {M.}~\bibnamefont {Paúr}}, \bibinfo {author} {\bibfnamefont {B.}~\bibnamefont {Stoklasa}}, \bibinfo {author} {\bibfnamefont {Z.}~\bibnamefont {Hradil}}, \bibinfo {author} {\bibfnamefont {L.~L.}\ \bibnamefont {Sánchez-Soto}},\ and\ \bibinfo {author} {\bibfnamefont {J.}~\bibnamefont {Rehacek}},\ }\bibfield  {title} {\bibinfo {title} {Achieving the ultimate optical resolution},\ }\href {https://doi.org/10.1364/OPTICA.3.001144} {\bibfield  {journal} {\bibinfo  {journal} {Optica}\ }\textbf {\bibinfo {volume} {3}},\ \bibinfo {pages} {1144} (\bibinfo {year} {2016})}\BibitemShut {NoStop}%
\bibitem [{\citenamefont {Tham}\ \emph {et~al.}(2017)\citenamefont {Tham}, \citenamefont {Ferretti},\ and\ \citenamefont {Steinberg}}]{Tham_Ferretti_Steinberg_2017}%
  \BibitemOpen
  \bibfield  {author} {\bibinfo {author} {\bibfnamefont {W.-K.}\ \bibnamefont {Tham}}, \bibinfo {author} {\bibfnamefont {H.}~\bibnamefont {Ferretti}},\ and\ \bibinfo {author} {\bibfnamefont {A.~M.}\ \bibnamefont {Steinberg}},\ }\bibfield  {title} {\bibinfo {title} {Beating rayleigh’s curse by imaging using phase information},\ }\href {https://doi.org/10.1103/PhysRevLett.118.070801} {\bibfield  {journal} {\bibinfo  {journal} {Physical Review Letters}\ }\textbf {\bibinfo {volume} {118}},\ \bibinfo {pages} {070801} (\bibinfo {year} {2017})}\BibitemShut {NoStop}%
\bibitem [{\citenamefont {Paúr}\ \emph {et~al.}(2018)\citenamefont {Paúr}, \citenamefont {Stoklasa}, \citenamefont {Grover}, \citenamefont {Krzic}, \citenamefont {Sánchez-Soto}, \citenamefont {Hradil},\ and\ \citenamefont {Řeháček}}]{Paur_Stoklasa_Grover_Krzic_Sanchez-Soto_Hradil_Rehacek_2018}%
  \BibitemOpen
  \bibfield  {author} {\bibinfo {author} {\bibfnamefont {M.}~\bibnamefont {Paúr}}, \bibinfo {author} {\bibfnamefont {B.}~\bibnamefont {Stoklasa}}, \bibinfo {author} {\bibfnamefont {J.}~\bibnamefont {Grover}}, \bibinfo {author} {\bibfnamefont {A.}~\bibnamefont {Krzic}}, \bibinfo {author} {\bibfnamefont {L.~L.}\ \bibnamefont {Sánchez-Soto}}, \bibinfo {author} {\bibfnamefont {Z.}~\bibnamefont {Hradil}},\ and\ \bibinfo {author} {\bibfnamefont {J.}~\bibnamefont {Řeháček}},\ }\bibfield  {title} {\bibinfo {title} {Tempering rayleigh’s curse with psf shaping},\ }\href {https://doi.org/10.1364/optica.5.001177} {\bibfield  {journal} {\bibinfo  {journal} {Optica}\ }\textbf {\bibinfo {volume} {5}},\ \bibinfo {pages} {1177} (\bibinfo {year} {2018})}\BibitemShut {NoStop}%
\bibitem [{\citenamefont {Donohue}\ \emph {et~al.}(2018)\citenamefont {Donohue}, \citenamefont {Ansari}, \citenamefont {Řeh{\'{a}}{\v{c}}ek}, \citenamefont {Hradil}, \citenamefont {Stoklasa}, \citenamefont {Pa{\'{u}}r}, \citenamefont {S{\'{a}}nchez-Soto},\ and\ \citenamefont {Silberhorn}}]{Donohue2018}%
  \BibitemOpen
  \bibfield  {author} {\bibinfo {author} {\bibfnamefont {J.~M.}\ \bibnamefont {Donohue}}, \bibinfo {author} {\bibfnamefont {V.}~\bibnamefont {Ansari}}, \bibinfo {author} {\bibfnamefont {J.}~\bibnamefont {Řeh{\'{a}}{\v{c}}ek}}, \bibinfo {author} {\bibfnamefont {Z.}~\bibnamefont {Hradil}}, \bibinfo {author} {\bibfnamefont {B.}~\bibnamefont {Stoklasa}}, \bibinfo {author} {\bibfnamefont {M.}~\bibnamefont {Pa{\'{u}}r}}, \bibinfo {author} {\bibfnamefont {L.~L.}\ \bibnamefont {S{\'{a}}nchez-Soto}},\ and\ \bibinfo {author} {\bibfnamefont {C.}~\bibnamefont {Silberhorn}},\ }\bibfield  {title} {\bibinfo {title} {{Quantum-Limited Time-Frequency Estimation through Mode-Selective Photon Measurement}},\ }\href {https://doi.org/10.1103/PhysRevLett.121.090501} {\bibfield  {journal} {\bibinfo  {journal} {Physical Review Letters}\ }\textbf {\bibinfo {volume} {121}},\ \bibinfo {pages} {1} (\bibinfo {year} {2018})},\ \Eprint {https://arxiv.org/abs/1805.02491} {arXiv:1805.02491} \BibitemShut {NoStop}%
\bibitem [{\citenamefont {Boucher}\ \emph {et~al.}(2020)\citenamefont {Boucher}, \citenamefont {Fabre}, \citenamefont {Labroille},\ and\ \citenamefont {Treps}}]{Boucher_Fabre_Labroille_Treps_2020}%
  \BibitemOpen
  \bibfield  {author} {\bibinfo {author} {\bibfnamefont {P.}~\bibnamefont {Boucher}}, \bibinfo {author} {\bibfnamefont {C.}~\bibnamefont {Fabre}}, \bibinfo {author} {\bibfnamefont {G.}~\bibnamefont {Labroille}},\ and\ \bibinfo {author} {\bibfnamefont {N.}~\bibnamefont {Treps}},\ }\bibfield  {title} {\bibinfo {title} {Spatial optical mode demultiplexing as a practical tool for optimal transverse distance estimation},\ }\href {https://doi.org/10.1364/optica.404746} {\bibfield  {journal} {\bibinfo  {journal} {Optica}\ }\textbf {\bibinfo {volume} {7}},\ \bibinfo {pages} {1621} (\bibinfo {year} {2020})}\BibitemShut {NoStop}%
\bibitem [{Sup()}]{Supplement}%
  \BibitemOpen
  \href@noop {} {\bibinfo  {journal} {See Supplemental Material for more information}\ }\BibitemShut {NoStop}%
\bibitem [{\citenamefont {Kendall}\ and\ \citenamefont {Stuart}(1961)}]{Kendall_Stuart_1961}%
  \BibitemOpen
\bibfield  {journal} {  }\bibfield  {author} {\bibinfo {author} {\bibfnamefont {M.~G.}\ \bibnamefont {Kendall}}\ and\ \bibinfo {author} {\bibfnamefont {A.}~\bibnamefont {Stuart}},\ }\href@noop {} {\emph {\bibinfo {title} {The Advanced Theory of Statistics, Vol. 2: Inference and Relationship}}}\ (\bibinfo  {publisher} {Charles Griffin and Company Ltd.},\ \bibinfo {year} {1961})\BibitemShut {NoStop}%
\bibitem [{\citenamefont {Cramér}(1999)}]{Cramer_1999}%
  \BibitemOpen
  \bibfield  {author} {\bibinfo {author} {\bibfnamefont {H.}~\bibnamefont {Cramér}},\ }\href@noop {} {\emph {\bibinfo {title} {Mathematical Methods of Statistics}}}\ (\bibinfo  {publisher} {Princeton University Press},\ \bibinfo {year} {1999})\BibitemShut {NoStop}%
\bibitem [{\citenamefont {Bisketzi}\ \emph {et~al.}(2019)\citenamefont {Bisketzi}, \citenamefont {Branford},\ and\ \citenamefont {Datta}}]{Bisketzi_Branford_Datta_2019}%
  \BibitemOpen
  \bibfield  {author} {\bibinfo {author} {\bibfnamefont {E.}~\bibnamefont {Bisketzi}}, \bibinfo {author} {\bibfnamefont {D.}~\bibnamefont {Branford}},\ and\ \bibinfo {author} {\bibfnamefont {A.}~\bibnamefont {Datta}},\ }\bibfield  {title} {\bibinfo {title} {Quantum limits of localisation microscopy},\ }\href {https://doi.org/10.1088/1367-2630/ab58a0} {\bibfield  {journal} {\bibinfo  {journal} {New Journal of Physics}\ }\textbf {\bibinfo {volume} {21}},\ \bibinfo {pages} {123032} (\bibinfo {year} {2019})}\BibitemShut {NoStop}%
\bibitem [{\citenamefont {Liu}\ \emph {et~al.}(2020)\citenamefont {Liu}, \citenamefont {Yuan}, \citenamefont {Lu},\ and\ \citenamefont {Wang}}]{Liu2020}%
  \BibitemOpen
  \bibfield  {author} {\bibinfo {author} {\bibfnamefont {J.}~\bibnamefont {Liu}}, \bibinfo {author} {\bibfnamefont {H.}~\bibnamefont {Yuan}}, \bibinfo {author} {\bibfnamefont {X.-M.}\ \bibnamefont {Lu}},\ and\ \bibinfo {author} {\bibfnamefont {X.}~\bibnamefont {Wang}},\ }\bibfield  {title} {\bibinfo {title} {Quantum fisher information matrix and multiparameter estimation},\ }\href {https://doi.org/10.1088/1751-8121/ab5d4d} {\bibfield  {journal} {\bibinfo  {journal} {Journal of Physics A: Mathematical and Theoretical}\ }\textbf {\bibinfo {volume} {53}},\ \bibinfo {pages} {023001} (\bibinfo {year} {2020})}\BibitemShut {NoStop}%
\bibitem [{\citenamefont {Peng}\ and\ \citenamefont {Lu}(2021)}]{Peng_Lu_2021}%
  \BibitemOpen
  \bibfield  {author} {\bibinfo {author} {\bibfnamefont {L.}~\bibnamefont {Peng}}\ and\ \bibinfo {author} {\bibfnamefont {X.~M.}\ \bibnamefont {Lu}},\ }\bibfield  {title} {\bibinfo {title} {Generalization of rayleigh’s criterion on parameter estimation with incoherent sources},\ }\href {https://doi.org/10.1103/PhysRevA.103.042601} {\bibfield  {journal} {\bibinfo  {journal} {Physical Review A}\ }\textbf {\bibinfo {volume} {103}},\ \bibinfo {pages} {1–10} (\bibinfo {year} {2021})}\BibitemShut {NoStop}%
\bibitem [{\citenamefont {Paúr}\ \emph {et~al.}(2019)\citenamefont {Paúr}, \citenamefont {Stoklasa}, \citenamefont {Koutný}, \citenamefont {Řeháček}, \citenamefont {Hradil}, \citenamefont {Grover}, \citenamefont {Krzic},\ and\ \citenamefont {Sánchez-Soto}}]{Paur_Stoklasa_Koutny_Rehacek_Hradil_Grover_Krzic_Sanchez-Soto_2019}%
  \BibitemOpen
  \bibfield  {author} {\bibinfo {author} {\bibfnamefont {M.}~\bibnamefont {Paúr}}, \bibinfo {author} {\bibfnamefont {B.}~\bibnamefont {Stoklasa}}, \bibinfo {author} {\bibfnamefont {D.}~\bibnamefont {Koutný}}, \bibinfo {author} {\bibfnamefont {J.}~\bibnamefont {Řeháček}}, \bibinfo {author} {\bibfnamefont {Z.}~\bibnamefont {Hradil}}, \bibinfo {author} {\bibfnamefont {J.}~\bibnamefont {Grover}}, \bibinfo {author} {\bibfnamefont {A.}~\bibnamefont {Krzic}},\ and\ \bibinfo {author} {\bibfnamefont {L.~L.}\ \bibnamefont {Sánchez-Soto}},\ }\bibfield  {title} {\bibinfo {title} {Reading out fisher information from the zeros of the point spread function},\ }\href {https://doi.org/10.1364/ol.44.003114} {\bibfield  {journal} {\bibinfo  {journal} {Optics Letters}\ }\textbf {\bibinfo {volume} {44}},\ \bibinfo {pages} {3114} (\bibinfo {year} {2019})}\BibitemShut {NoStop}%
\bibitem [{\citenamefont {Svensson}(2013)}]{svensson2013pedagogical}%
  \BibitemOpen
  \bibfield  {author} {\bibinfo {author} {\bibfnamefont {B.~E.}\ \bibnamefont {Svensson}},\ }\bibfield  {title} {\bibinfo {title} {Pedagogical review of quantum measurement theory with an emphasis on weak measurements},\ }\href@noop {} {\bibfield  {journal} {\bibinfo  {journal} {Quanta}\ }\textbf {\bibinfo {volume} {2}},\ \bibinfo {pages} {18} (\bibinfo {year} {2013})}\BibitemShut {NoStop}%
\bibitem [{\citenamefont {Struchalin}\ \emph {et~al.}(2018)\citenamefont {Struchalin}, \citenamefont {Kovlakov}, \citenamefont {Straupe},\ and\ \citenamefont {Kulik}}]{struchalin2018adaptive}%
  \BibitemOpen
  \bibfield  {author} {\bibinfo {author} {\bibfnamefont {G.}~\bibnamefont {Struchalin}}, \bibinfo {author} {\bibfnamefont {E.}~\bibnamefont {Kovlakov}}, \bibinfo {author} {\bibfnamefont {S.}~\bibnamefont {Straupe}},\ and\ \bibinfo {author} {\bibfnamefont {S.}~\bibnamefont {Kulik}},\ }\bibfield  {title} {\bibinfo {title} {Adaptive quantum tomography of high-dimensional bipartite systems},\ }\href@noop {} {\bibfield  {journal} {\bibinfo  {journal} {Physical Review A}\ }\textbf {\bibinfo {volume} {98}},\ \bibinfo {pages} {032330} (\bibinfo {year} {2018})}\BibitemShut {NoStop}%
\bibitem [{\citenamefont {Bogdanov}\ \emph {et~al.}(2020)\citenamefont {Bogdanov}, \citenamefont {Bolshedvorskii}, \citenamefont {Zeleneev}, \citenamefont {Soshenko}, \citenamefont {Rubinas}, \citenamefont {Radishev}, \citenamefont {Lobaev}, \citenamefont {Vikharev}, \citenamefont {Gorbachev}, \citenamefont {Drozdov} \emph {et~al.}}]{bogdanov2020optical}%
  \BibitemOpen
  \bibfield  {author} {\bibinfo {author} {\bibfnamefont {S.}~\bibnamefont {Bogdanov}}, \bibinfo {author} {\bibfnamefont {S.}~\bibnamefont {Bolshedvorskii}}, \bibinfo {author} {\bibfnamefont {A.}~\bibnamefont {Zeleneev}}, \bibinfo {author} {\bibfnamefont {V.}~\bibnamefont {Soshenko}}, \bibinfo {author} {\bibfnamefont {O.}~\bibnamefont {Rubinas}}, \bibinfo {author} {\bibfnamefont {D.}~\bibnamefont {Radishev}}, \bibinfo {author} {\bibfnamefont {M.}~\bibnamefont {Lobaev}}, \bibinfo {author} {\bibfnamefont {A.}~\bibnamefont {Vikharev}}, \bibinfo {author} {\bibfnamefont {A.}~\bibnamefont {Gorbachev}}, \bibinfo {author} {\bibfnamefont {M.}~\bibnamefont {Drozdov}}, \emph {et~al.},\ }\bibfield  {title} {\bibinfo {title} {Optical investigation of as-grown nv centers in heavily nitrogen doped delta layers in cvd diamond},\ }\href@noop {} {\bibfield  {journal} {\bibinfo  {journal} {Materials Today Communications}\ }\textbf {\bibinfo {volume} {24}},\ \bibinfo {pages} {101019} (\bibinfo {year} {2020})}\BibitemShut
  {NoStop}%
\bibitem [{\citenamefont {Santamaria}\ \emph {et~al.}(2023)\citenamefont {Santamaria}, \citenamefont {Pallotti}, \citenamefont {de~Cumis}, \citenamefont {Dequal},\ and\ \citenamefont {Lupo}}]{santamaria2023spatial}%
  \BibitemOpen
  \bibfield  {author} {\bibinfo {author} {\bibfnamefont {L.}~\bibnamefont {Santamaria}}, \bibinfo {author} {\bibfnamefont {D.}~\bibnamefont {Pallotti}}, \bibinfo {author} {\bibfnamefont {M.~S.}\ \bibnamefont {de~Cumis}}, \bibinfo {author} {\bibfnamefont {D.}~\bibnamefont {Dequal}},\ and\ \bibinfo {author} {\bibfnamefont {C.}~\bibnamefont {Lupo}},\ }\bibfield  {title} {\bibinfo {title} {Spatial-mode demultiplexing for enhanced intensity and distance measurement},\ }\href@noop {} {\bibfield  {journal} {\bibinfo  {journal} {Optics Express}\ }\textbf {\bibinfo {volume} {31}},\ \bibinfo {pages} {33930} (\bibinfo {year} {2023})}\BibitemShut {NoStop}%
\bibitem [{\citenamefont {Richter}\ \emph {et~al.}(2021)\citenamefont {Richter}, \citenamefont {Wolf}, \citenamefont {Von~Zanthier},\ and\ \citenamefont {{Schmidt-Kaler}}}]{richter_imaging_2021}%
  \BibitemOpen
  \bibfield  {author} {\bibinfo {author} {\bibfnamefont {S.}~\bibnamefont {Richter}}, \bibinfo {author} {\bibfnamefont {S.}~\bibnamefont {Wolf}}, \bibinfo {author} {\bibfnamefont {J.}~\bibnamefont {Von~Zanthier}},\ and\ \bibinfo {author} {\bibfnamefont {F.}~\bibnamefont {{Schmidt-Kaler}}},\ }\bibfield  {title} {\bibinfo {title} {Imaging {{Trapped Ion Structures}} via {{Fluorescence Cross-Correlation Detection}}},\ }\href {https://doi.org/10.1103/PhysRevLett.126.173602} {\bibfield  {journal} {\bibinfo  {journal} {Physical Review Letters}\ }\textbf {\bibinfo {volume} {126}},\ \bibinfo {pages} {173602} (\bibinfo {year} {2021})}\BibitemShut {NoStop}%
\end{thebibliography}%


\begin{thebibliography}{1}

\bibitem{Bisketzi_Branford_Datta_2019}
Evangelia Bisketzi, Dominic Branford, and Animesh Datta.
\newblock Quantum limits of localisation microscopy.
\newblock {\em New Journal of Physics}, 21(12):123032, Dec 2019.

\bibitem{Cramer_1999}
Harald Cramér.
\newblock {\em Mathematical Methods of Statistics}.
\newblock Princeton University Press, 1999.

\bibitem{Kendall_Stuart_1961}
M.~G. Kendall and A.~Stuart.
\newblock {\em The Advanced Theory of Statistics, Vol. 2: Inference and Relationship}.
\newblock Charles Griffin and Company Ltd., 1961.

\bibitem{Liu2020}
Jing Liu, Haidong Yuan, Xiao-Ming Lu, and Xiaoguang Wang.
\newblock Quantum fisher information matrix and multiparameter estimation.
\newblock {\em Journal of Physics A: Mathematical and Theoretical}, 53:023001, 1 2020.

\bibitem{Tsang_Nair_Lu_2015}
Mankei Tsang, Ranjith Nair, and Xiao-Ming Lu.
\newblock Quantum theory of superresolution for two incoherent optical point sources.
\newblock {\em Physical Review X}, 6(3):031033, 2016.

\end{thebibliography}

\end{document}


\maketitle
\tableofcontents

\newpage
\section{Statistical error propagation theory}\label{sec:Errors}

\subsection{Mean photon number measurement in $\HG{1}$ mode (SPADE and BLESS protocols)}
\label{ss:HG1}

Consider the measurement of the mean photon number in the $\HG{1}$ detection mode, centered at the point $x_D=x_c$. This measurement corresponds to the second stage of the SPADE and BLESS protocols, as described in the main text and in Sec.~\ref{sec:protocols}. In the small distance approximation $(d \ll \sigma)$, the mean photon number can be approximated as (7,9):
\begin{equation}\label{eq:Mxc}
\mu_{x_c}\equiv \mu_1(x_c)\approx\frac{\mu g d^2}{2\sigma^2},\quad
g=g^{(2)}(t=0)=\frac{1-\gamma^2}{2}.
\end{equation}
Thus, the distance $d$ can be estimated as:
\begin{equation}\label{eq:dMxc}
    d = \sigma \sqrt{\frac{2\mu_{x_c}}{\mu g}}.
\end{equation}

Assuming that $g$ and $\mu$ are accurately estimated, the distance estimation error $\Delta_d$ is determined by the statistical errors in the measured mean photon number $\mu_{x_c}$. For $\mu \ll 1$, we can approximate the mean photon number as $\mu_{x_c} \approx P_{1x_c} \equiv P_1(1|x_D=x_c)$. With a large number of measurements $N \gg 1$, the error in $\mu_{x_c}$ is:
\begin{equation}\label{eq:DeltaMxc}
    \Delta_{\mu_{x_c}}^2 = \Delta_{P_{1x_c}}^2 = \frac{P_{1x_c}}{N} = \frac{\mu_{x_c}}{N}.
\end{equation}
Here, $P_1(n)$ represents the total detected photon number distribution (PND) in the $\HG{1}$ detection mode (5).

Finally, by using \eqref{eq:DeltaMxc} and \eqref{eq:Mxc}, the distance estimation error is calculated as:
\begin{equation}\label{eq:DeltadMxc}
\Delta_d = \left| \frac{\partial d}{\partial \mu_{x_c}} \right| \Delta_{\mu_{x_c}}
= \frac{d}{2 \mu_{x_c}} \Delta_{\mu_{x_c}} = \frac{\sigma}{\sqrt{2\mu g N}},
\end{equation}
which is consistent with Eq.~(8) in the main text.

\subsection{Photon number distribution measurement in $\HG{0}$ mode (ESS protocol)}\label{ss:HG0}

Consider the PND measurement in the $\HG{0}$ mode, centered at the points $x_D=\pm\sigma$, assuming $x_c \ll \sigma$ and $d \ll \sigma$. This corresponds to the ESS protocol described in the main text and in Sec.~\ref{sec:protocols}, with the mesh reduced to two points $\mathbf{X_0} = \{-1, +1\}\sigma$.

The probabilities of detecting two photons (Eq.~5) are given by:
\begin{equation}\label{eq:P2}
    P_{2\pm} \equiv P_0(2|x_D=\pm\sigma) = \mu^2 g \frac{1 \pm 2 x_c/\sigma \mp \gamma d/\sigma}{2 \text{e}},
    \quad g = g^{(2)}(t=0) = \frac{1 - \gamma^2}{2}.
\end{equation}
From these probabilities, the parameters $g$ and $\gamma$ can be estimated as:
\begin{equation}\label{eq:gP2}
    g = 2 \text{e} \frac{P_{2+} + P_{2-}}{\mu^2}, \quad \gamma = \pm\sqrt{1 - 2g}.
\end{equation}
The distance $d$ can then be estimated using:
\begin{equation}\label{eq:dP2}
    d = 2\frac{x_c}{\gamma} - \frac{\text{e} \sigma}{\mu^2 g \gamma}(P_{2+} - P_{2-}).
\end{equation}

Assuming accurate estimates of $x_c$, $\gamma$, and $\mu$, the distance estimation error $\Delta_d$ is determined by the statistical errors in the measured probabilities $P_{2\pm}$. For a large number of measurements ($N \gg 1$), these errors are given by:
\begin{equation}\label{eq:DeltaP2}
    \Delta_{P_{2\pm}}^2 = \frac{P_{2\pm}}{N} \approx \frac{\mu^2 g}{2 \text{e} N}.
\end{equation}

Finally, the distance estimation error can be calculated as:
\begin{equation}\label{eq:DeltadP2}
    \Delta_d = \sqrt{\left(\frac{\partial d}{\partial P_{2+}}\right)^2 \Delta_{P_{2+}}^2
    + \left(\frac{\partial d}{\partial P_{2-}}\right)^2 \Delta_{P_{2-}}^2}
    \approx \sqrt{2} \text{e} \sigma \frac{\Delta_{P_{2\pm}}}{\mu^2 g \gamma}
    \approx \sqrt{\frac{\text{e} \sigma^2}{\mu^2 \gamma^2 g N}},
\end{equation}
which is equivalent to Eq.~(10) in the main text.

\newpage
\section{Classical Fisher information and Cram{\'e}r--Rao bound calculation}\label{sec:FIM}

\subsection{Fisher information and Cram{\'e}r--Rao bound definition}

To calculate the parameters' estimation errors for different measurement protocols, we use the Cram{\'e}r--Rao bound (CRB) theory \cite{Cramer_1999,Kendall_Stuart_1961}. Consider the vector of estimated object parameters $\boldsymbol{\theta} = \{d, \gamma, x_c, \mu\}$ and its unbiased estimate $\hat{\boldsymbol{\theta}}$. For each imaging experiment, $\hat{\boldsymbol{\theta}}$ is a random vector centered at the true parameter values $\boldsymbol{\theta}^\ast$. 

In the limit of an infinitely large number of measurements ($N \rightarrow \infty$), the estimate vector $\hat{\boldsymbol{\theta}}$ follows a multivariate normal distribution with the covariance matrix $\Sigma$. The diagonal elements of this matrix represent the variances of the estimated parameters: $\Delta_d^2$, $\Delta_\gamma^2$, $\Delta_{x_c}^2$, and $\Delta_\mu^2$. According to CRB, the covariance matrix $\Sigma$ is bounded by the inverse of the positive semidefinite Fisher information matrix (FIM) $F$ multiplied by the number of measurements $N$. Since FIM is additive with respect to independent measurements, we sum up FIMs correspondent to different detector positions and measurement regimes.

In the \textbf{photon number distribution (PND) measurement regime}, the FIM is given by:
\begin{equation}
    F^{\text{PND}}_m(x_D) = \sum_n \frac{\left[\nabla_{\boldsymbol{\theta}} P_m(n)\right]\left[\nabla_{\boldsymbol{\theta}} P_m(n)\right]^T}{P_m(n)},
\end{equation}
where $P_m(n)$ is the detected PND (Eq.~5) that depends on $\boldsymbol{\theta}$ and the known detection mode parameters: position $x_D$ and the mode index $\HG{m}$.

In the \textbf{mean photon number measurement regime}, the FIM is computed based on the measured mean photon number, with an expected value $\mu_m=\mu_{am}+\mu_{bm}$ and variance $V_m = \mu_{am}(1 - \mu_{am}) + \mu_{bm}(1 - \mu_{bm})$. Assuming normal statistical errors, the FIM is given by: \begin{equation} F^{\text{mean}}_m(x_D) = \frac{\left[\nabla_{\boldsymbol{\theta}} \mu_m\right] \left[\nabla_{\boldsymbol{\theta}} \mu_m\right]^T}{V_m}. \end{equation}

\subsection{Fisher information matrix for different measurement protocols}\label{sec:protocols}

Since the FIM is additive with respect to independent measurements, we define $F_\text{mesh} = \frac{1}{11}\sum_{x_D \in \mathbf{X_0}}{F_0(x_D)}$, where $\mathbf{X_0} = \{-1, -0.8, -0.6, \dots, +0.8, +1\} \times \sigma$ represents the 11-point mesh scanning the sample in the $\HG{0}$ mode.

As shown in Sec.~\ref{ss:HG0}, measuring at just two points $\mathbf{X_0} = \{-1,  +1\}\times \sigma$ is sufficient to estimate all system parameters using ESS and BLESS protocols. However, we increased the number of mesh points
to accommodate the DI and SPADE protocols as well.  Since our system is described by four parameters, the mesh requires at least four points. We increased the number of points for smoother results, but, as we found later, the final precision bounds do not significantly depend on the number of mesh points.

\textbf{Direct imaging (DI)} aims to extract information from intensity measurements in the standard $\HG{0}$ mode. Hence, $F_{\text{DI}} = N F_\text{mesh}^{\text{mean}}$, where $N$ is the total number of shots (proportional to the total exposure time).

After determining the centroid location $\tilde{x}_c$ through direct imaging, the \textbf{SPADE protocol} measures the mean photon number at $x_D = \tilde{x}_c$ in the $\HG{1}$ mode. 
To simplify analysis, we assume the estimated centroid location $\tilde{x}_c$ has an error small enough to make the first term in the mean photon number approximation around $x_c$:
\begin{equation}\label{eq:parabola}
\mu_1(x_D) \approx \frac{\mu}{2\sigma^2}\left[(x_D - x_c)^2 + g d^2\right],
\end{equation}
negligible. 
Specifically, we choose $\tilde{x}_c = x_c + d\sqrt{g/10}$, so that the first term in \eqref{eq:parabola} is ten times smaller than the second one. Assuming both stages of the protocol take $N/2$ measurement samples, we then have $F_{\text{SPADE}} = \frac{N}{2} \left[F_\text{mesh}^{\text{mean}} + F_1^{\text{mean}}(\tilde{x}_c)\right]$.


Due to the strong correlation between the relative brightness $\gamma$ and the distance $d$, these measurements alone are insufficient for precise distance estimation. In the main text, we propose measuring the photon number distribution. \textbf{Estimating the shot statistics (ESS)} at points $\mathbf{X_0}$ gives $F_{\text{ESS}} = N F_\text{mesh}^{\text{PND}}$. While this measurement alone breaks Rayleigh’s curse -- making the absolute error independent of $d$ -- the resulting error remains relatively high.

To improve precision, we extend the ESS protocol to the \textbf{BLESS protocol} by adding an additional mean photon number measurement in the modulated $\HG{1}$ detection mode, yielding $F_{\text{BLESS}} = \frac{N}{2} \left(F_\text{mesh}^{\text{PND}} + F_1^{\text{mean}}(\tilde{x}_c)\right)$.

\newpage
\section{Experimental imperfections: cross-talk and background noise}\label{sec:Noise}
\subsection{Mathematical model of cross-talk and background noise}\label{ss:NoiseModel}

We address two main experimental imperfections. First, non-ideal spatial modulation caused by imperfections in the spatial light modulator and optical aberrations lead to cross-talk between the \Lab{HG_0} and \Lab{HG_1} modes. This introduces a non-zero coupling coefficient $\left.\Tilde{C}_{s1}\right|_{x_D=x_s}=\chi$, which we model as $\Tilde{C}_{s1}=\chi C_{s0} + (1-\chi) C_{s1}$.

Second, we consider background radiation modeled by a Poissonian photon number distribution (PND), $P_\text{bg}(n)=\frac{\mu_{\text{bg}}^n}{n!}{\rm e}^{-\mu_{\text{bg}}}$, characterized by the mean photon number $\mu_\text{bg}$. The total probability of detecting $n$ photons, $\Tilde{P}_m(n)$, is calculated by convolving $P_\text{bg}(n)$ with ${P}_m(n)$. Assuming the photon number resolving detector (PNRD) resolves up to two photons, we sum the probabilities $\Tilde{P}_m(n\geq 2)$.

\subsection{Estimation the impact of imperfections, based on the statistical error propagation theory}

\subsubsection{Impact of imperfections on the BLESS protocol}\label{ss:HG1noise}

Similar to Sec.~\ref{ss:HG1}, consider the measurement of the mean photon number in the $\HG{1}$ mode centered at $x_D=x_c$. In the small distance approximation $(d\ll\sigma)$, the detected mean photon number (7) is modified to account for background noise with mean photon number $\mu_\text{bg}$ and spatial mode cross-talk $\chi$:

\begin{equation}\label{eq:Mxc1}
\mu_{x_c}\equiv \mu_1(x_c)\approx\frac{\mu g d^2}{2\sigma^2}+\mu_\text{bg}+\mu\chi,
    \quad
g=g^{(2)}(t=0)=\frac{1-\gamma^2}{2}.
\end{equation}
Thus, the distance $d$ can be estimated as:
\begin{equation}\label{eq:dMxc1}
    d=\sigma\sqrt{2\frac{ \mu_{x_c}-\mu_\text{bg}-\mu\chi}{\mu g}}.
\end{equation}

Assuming accurate estimates of $g$ and $\mu$, we can use Eq.~\eqref{eq:DeltaMxc} for $\Delta_{\mu_{x_c}}$ and Eq.~\eqref{eq:Mxc1} for $\mu_{x_c}$ to estimate the distance error as:

\begin{equation}\label{eq:DeltadMxc1}
\Delta_d=\abs{\frac{\partial d}{\partial \mu_{x_c}}}\Delta_{\mu_{x_c}}
=\frac{\sigma^2}{\mu g d}\sqrt{\frac{\mu_{x_c}}{N}}
=\frac{\sigma}{\mu g d}\sqrt{\frac{\mu g d^2+2\sigma^2[\mu_\text{bg}+\mu\chi]}{2N}}.
\end{equation}

This equation shows that for $d\gg d_c$, where 
\begin{equation}\label{eq:dc}
    d_c\equiv \sigma \sqrt{2\frac{\mu_\text{bg}+\mu\chi}{\mu g}},
\end{equation}
the distance estimation error $\Delta_d$ reduces to the initial case \eqref{eq:DeltadMxc}. However, for $d\ll d_c$, the error increases as $\Delta_d\propto d^{-1}$:

\begin{equation}\label{eq:DeltadMxc1}
\Delta_d\approx
\frac{\sigma^2}{\mu g d}\sqrt{\frac{\mu_\text{bg}+\mu\chi}{N}},
\end{equation}
indicating the return of Rayleigh's curse.
This behavior matches the CRB plots presented in Figs.~3(g) and \ref{fig:background}~(b,c).

\subsubsection{
Impact of imperfections on the BLESS protocol}\label{ss:HG0noise}

Similar to Sec.~\ref{ss:HG1}, considering measurements in the $\HG{0}$ mode, we can neglect cross-talk, but background noise significantly alters the photon statistics. As mentioned above, the total probability to detect $n$ photons, $\Tilde{P}_m(n)$, is the convolution of ${P}_m(n)$ with the Poisson distribution $P_\text{bg}(n)$. Thus, Eq.~(5) for the probability distribution $P_m(n|x_D)$ is modified as follows:

\begin{subequations}\label{eq.PdPbg}
    \begin{align}
    \Tilde{P}_m(0)=&P_\text{bg}(0){P}_m(0),\\
    \Tilde{P}_m(1)=&P_\text{bg}(1){P}_m(0),
    +P_\text{bg}(0){P}_m(1)\\
    \Tilde{P}_m(2)=&1-\Tilde{P}_m(0)-\Tilde{P}_m(1).
    \end{align}
\end{subequations}

For measurements in the $\HG{0}$ mode centered at the points $x_D=\pm\sigma$, the probabilities to detect two photons, under the assumptions $d\ll\sigma$, $x_c\ll\sigma$, and $\mu_\text{bg}\ll 1$, are:

\begin{equation}
    P_{2\pm}\equiv \Tilde{P}_0(2|x_D=\pm\sigma)
    =\mu\frac{ \mu g (1\pm 2 x_c d/ \sigma \mp \gamma d/ \sigma) + 2 \sqrt{\text{e}}\mu_\text{bg}  }{2 \text{e}}
\end{equation}
Thus, the parameters $g$ and $\gamma$ can be estimated as:
\begin{equation}\label{eq:gnoise}
    g=\sqrt{\text{e}} \frac{\sqrt{\text{e}}(P_{2+}+P_{2-})-2\mu \mu_\text{bg}}{\mu^2}, \quad \gamma=\pm\sqrt{1-2g},
\end{equation}
and the distance $d$ can be estimated as: 
\begin{equation}\label{eq:dnoise}
    d=2\frac{x_c}{\gamma}-\frac{\text{e}\sigma}{\mu^2 g \gamma}(P_{2+}-P_{2-}).
\end{equation}

Since the equations \eqref{eq:dnoise} and \eqref{eq:dP2} are the same, the background noise does not directly affect the precision of the $d$ estimation. However, it decreases the precision of the $g$ estimation due to the difference between \eqref{eq:gnoise} and \eqref{eq:gP2}, which indirectly slightly increases $\Delta_d$, as shown in Figs~3~(g) and \ref{fig:background}~(a).

\subsection{Cram{\'e}r--Rao bound calculation}\label{ss:CRBnoise}

The Cram{\'e}r--Rao bounds for the ESS and BLESS protocols for different values of $\mu_\text{bg}$ and $\chi$ were calculated as described in Sec.~\ref{sec:FIM}, taking into account the detection mode and detected PND modifications outlined in Sec.~\ref{ss:NoiseModel}.

\begin{figure}[hbt]
    \centering
    \includegraphics[width=\textwidth]{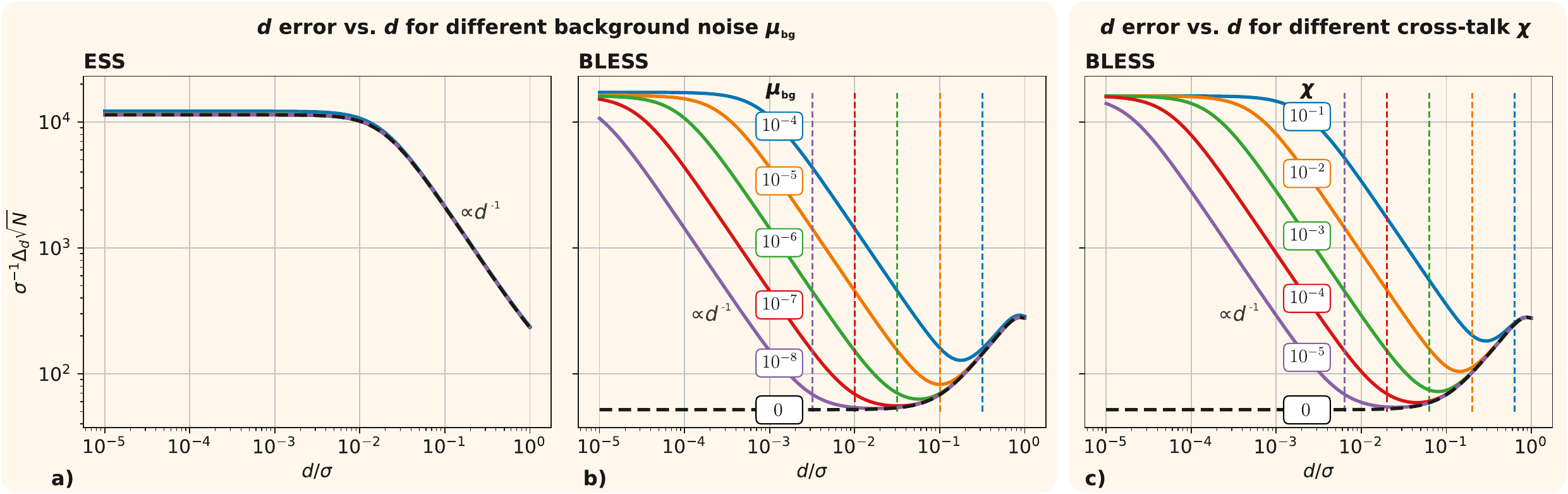}
    \caption{Cram{\'e}r--Rao bounds $\Delta_d$ vs. $d$ for ESS (a) and BLESS (b) protocols for different values of background noise $\mu_\text{bg}$ and for the BLESS protocol for different values of cross-talk $\chi$ (c).
    Corresponding values of $d_c$ \eqref{eq:dc} are plotted as vertical dashed lines in plots (b) and (c). In plot (a), different $\mu_\text{bg}$ values correspond to the same colors as in plot (b), but all the curves nearly overlap. All the plots were calculated with the following object parameters: $\gamma=0.1$, $x_c=0.25$, $\mu=4\times 10^{-3}$.}
    \label{fig:background}
\end{figure}

The results are presented in Fig.~\ref{fig:background}. 
Since the ESS protocol does not require measurements in the $\HG{1}$ mode, it is robust against cross-talk $\chi$, which defines the imperfection in detection mode shaping. However, as mentioned in Sec.~\ref{ss:HG0noise}, the ESS protocol can be disturbed by background noise $\mu_\text{bg}$. Nonetheless, Fig.~\ref{fig:background}~(a) shows that it remains quite robust to this imperfection, yielding $\Delta_d^\text{(ESS)}(d\rightarrow 0)\sim \sqrt{\frac{\sigma^2}{\mu^2\gamma^2 g N}}$ for all $\mu_\text{bg}$ values.

The BLESS protocol combines measurements in both $\HG{0}$ and $\HG{1}$ modes, so it is affected by both imperfections. As shown in Fig.~\ref{fig:background}~(b,c), for $d\gg d_c$, the estimation error $\Delta_d^\text{(BLESS)}$ is largely unaffected by cross-talk or background noise. However, for $d\ll d_c$, the error increases as $\Delta_d^\text{(BLESS)}\sim \frac{\sigma^2}{\mu g d}\sqrt{\frac{\mu_\text{bg}+\mu\chi}{N}}$ until it reaches $\Delta_d^\text{(BLESS)}(d\rightarrow 0)\sim\Delta_d^\text{(ESS)}(d\rightarrow 0)\sim \sqrt{\frac{\sigma^2}{\mu^2\gamma^2 g N}}$. Since in the BLESS protocol, half of the measurements are made in the $\HG{1}$ mode, which adds few information about $d$ at high error rates, the bound $\Delta_d^\text{(BLESS)}(d\rightarrow 0)$ is higher than $\Delta_d^\text{(ESS)}(d\rightarrow 0)$ by a factor of $\sqrt{2}$.

\newpage
\section{Quantum model of imaging}\label{sec:Quantum}

\subsection{Multiple-frequency quantum state of light}\label{SS:ImageDensityOperator}

Consider a single-photon source $s$ emitting a photon in the frequency mode $\omega$ with probability $\mu_s$:
\begin{equation}\label{eq:rho_s}
    \hat\rho_s^{(\omega)} = (1 - \mu_s)\ketbra{vac} + \mu_s \ketbra{\Psi_s}_\omega,
\end{equation}
where $\ket{\Psi_s}_\omega = \hat{A}_{s,\omega}^\dagger \ket{vac}$, and
\begin{equation}
    \hat{A}_{s,\omega}^\dagger = \int{{\rm d}q \Psi_s(q) \hat a^\dagger(q,\omega)}.
\end{equation}
We assume that the source emits a photon randomly into one of a large number of $M$ different frequency modes. Thus, the total density operator is a mixture of these modes: $\hat\rho_s=\frac{1}{M}\sum_{\omega}{\hat\rho_s^{(\omega)}}$. For example, in the case of NV-centers in diamond, photons are emitted together with a random number of phonons, and each phonon line is thermally broadened. The total number of spectral modes $M$ can be estimated as the product of spectral bandwidth and luminescence lifetime, which for NV-centers is approximately $M\approx 6\times 10^5$.

For two sources $a$ and $b$, the total density operator becomes:
\begin{equation}\label{eq:rho}
\begin{split}
    \hat\rho_{M}&=(1-\mu_a)(1-\mu_b)\ket{vac}\bra{vac}\\
    &+\frac{\mu_a(1-\mu_b)}{M}\sum\limits_{\omega_a} \hat A_{a,\omega_a}^\dagger \ket{vac}\bra{vac} \hat A_{a,\omega_a}\\
    &+\frac{\mu_b(1-\mu_a)}{M}\sum\limits_{\omega_b} \hat A_{b,\omega_b}^\dagger \ket{vac}\bra{vac} \hat A_{b,\omega_b}\\
    &+\frac{\mu_a \mu_b}{M^2}\sum\limits_{\omega_a, \omega_b} \frac{1}{N_{\omega_a \omega_b}} \hat A_{a,\omega_a}^\dagger \hat A_{b,\omega_b}^\dagger \ket{vac}\bra{vac} \hat A_{a,\omega_a}\hat A_{b,\omega_b}.
\end{split}
\end{equation}
Here, $N_{\omega_a \omega_b} = 1 + \abs{V}^2\delta_{\omega_a,\omega_b}$ is the normalization factor, and $V=\braket{\Psi_a}{\Psi_b}=\int{{\rm d}q \Psi_a^\ast(q) \Psi_b(q)}$ represents the overlap between the spatial modes of the two sources. When $M\gg 1$, the probability of both sources emitting photons in the same frequency mode ($\omega_a=\omega_b$) becomes negligible. Therefore, we simplify the state as:
\begin{equation}\label{eq:rho1}
    \hat\rho_M=\frac{1}{M(M-1)} \sum_{\omega_a\neq\omega_b}{\hat\rho_a^{(\omega_a)}\hat\rho_b^{(\omega_b)}} =\frac{1}{M(M-1)} \sum_{\omega_2>\omega_1}{\qty[\hat\rho_a^{(\omega_1)}\hat\rho_b^{(\omega_2)}+\hat\rho_b^{(\omega_1)}\hat\rho_a^{(\omega_2)}]}.
\end{equation}
An important property of this state is that it prevents the measurement of individual source parameters by filtering out particular modes.

\subsection{Photon number measurements}\label{ss:PhotonNumberMeasurements}

Let's consider the single-source state $\hat\rho_s^{(\omega)}$ \eqref{eq:rho_s} measured with a PNRD in the detection mode $\Phi(q)\equiv\Phi_0(q)$ at frequency $\omega$, where $\Phi_m(q)$ represents a set of orthogonal modes (e.g., $\HG{m}$ modes, $m=0,1,2,\dots$). For each mode $m$, we can introduce the Fock state basis $\{\ket{n_m}_{m,\omega}\}$, where $n_m$ is the number of photons in the mode $m$. Since the detector measures only the $m=0$ mode, we take the partial trace over the rest of the modes:
\begin{equation}
    \hat\rho_s^{(\omega,\Phi)} = \!\!\!\!\!\!\!\sum_{n_1,n_2,\dots=0}^{\infty}{\!\!\!\!\!\!\!\expval{\hat\rho_s^{(\omega)}}{\{n_{m>0}\}}_{m>0,\omega}}, \qquad \ket{\{n_{m>0}\}}_{m>0,\omega}=\Motimes_{m=1}^{\infty}{\ket{n_m}_{m,\omega}}.
\end{equation}
Since the state $\hat\rho_s^{(\omega)}$ contains at most one photon, this sum can be simplified as:
\begin{equation}
    \hat\rho_s^{(\omega,\Phi)} = \expval{\hat\rho_s^{(\omega)}}{0}_{m>0,\omega} + \sum_{m>0}{\expval{\hat\rho_s^{(\omega)}}{\tilde{1}_m}_{\omega}},
\end{equation}
where $\ket{\tilde{1}_m}_\omega$ is the state with one photon in the $m$ mode and zero photons in the other $m>0$ modes:
\begin{equation}
  \ket{\tilde{1}_m}_\omega=\qty[\Motimes_{m^\prime=1}^{m-1}{\ket{0}_{m^\prime,\omega}}]\ket{1}_{m,\omega}\qty[\Motimes_{m^\prime=m+1}^{\infty}{\ket{0}_{m^\prime,\omega}}].
\end{equation}
Define the coupling coefficient $C_{sm}=\abs*{\braket{\Psi_s}{1}_{m,\omega}}^2=\abs*{\int \Psi_s^\ast(q) \Phi_{m}(q){\rm d}q }^2$. We then find that:
\begin{equation}
    \expval{\hat\rho_s^{(\omega)}}{0}_{m>0,\omega}=(1-\mu_s)\ketbra{0}_{0,\omega} + \mu_s C_{s0}\ketbra{1}_{0,\omega}
\end{equation}
and
\begin{equation}
    \expval{\hat\rho_s^{(\omega)}}{\tilde{1}_m}_{\omega}=\mu_s C_{s0}\ketbra{0}_{0,\omega}.
\end{equation}
Since $\sum_{m>0}{C_{sm}}=1-C_{s0}$, after some simplifications, the detection mode density operator becomes:
\begin{equation}
    \hat\rho_{s0}^{(\omega)} = (1-\mu_s C_{s0})\ketbra{0}_\omega + \mu_s C_{s0}\ketbra{1}_\omega
\end{equation}

This procedure is performed for each frequency mode $\omega$. Then, similarly to \eqref{eq:rho1}, the density operator for two single-photon sources in the detection mode $\Phi_0$ is:

\begin{equation}
    \hat\rho_{M0}=\frac{1}{M(M-1)} \sum_{\omega_2>\omega_1}{\qty[\hat\rho_{a0}^{(\omega_1)}\hat\rho_{b0}^{(\omega_2)}+\hat\rho_{b0}^{(\omega_1)}\hat\rho_{a0}^{(\omega_2)}]}.
\end{equation}

Consider a PNRD that does not resolve different frequency modes. Such a PNRD would have the following POVM effects:

\begin{equation}
    \hat{E}_0^{(0)} = \ketbra{0}, \quad
    \hat{E}_0^{(1)} = \sum_{\omega}{\ketbra{1}_\omega}, \quad
    \hat{E}_0^{(2)} = \hat I - \hat{E}_0^{(0)} - \hat{E}_0^{(1)}.
\end{equation}

It can be easily verified that $P_0(n)=\Tr(\hat{E}_0^{(n)} \hat\rho_{M0})$ equals the classical distribution given by Eq.~(4) in the main text. 
Similarly, we can derive the corresponding equation for $P_1(n)$.

\subsection{Two-mode quantum state of light}\label{ss:TwoModeState}

The multi-mode state $\hat\rho_M$ \eqref{eq:rho1} can be reduced to a two-mode state as follows. Consider an arbitrary POVM $\{\hat{E}^{(n)}: \hat{E}^{(n)}\geq 0, \sum_n{\hat{E}^{(n)}}=\hat I\}$. The corresponding probabilities are:
\begin{equation}\label{eq:dm_pk1}
    p^{(n)}
    =\Tr(\hat{E}^{(n)}\hat\rho_M)
    =\frac{1}{M(M-1)} \sum_{\omega_2>\omega_1}{\Tr(E^{(n)}\qty[\hat\rho_a^{(\omega_1)}\hat\rho_b^{(\omega_2)}+\hat\rho_b^{(\omega_1)}\hat\rho_a^{(\omega_2)}])}.
\end{equation}

For each term on the right-hand side, let's first take the trace over the modes with $\omega\neq\omega_1,\omega_2$. Since the states in these modes are vacuum states, one can introduce the operator:

\begin{equation}\label{eq:povm_w1_w2}
    \hat{E}_{\omega_1,\omega_2}^{(n, \omega_1,\omega_2)} = \qty[\Motimes_{\omega\neq\omega_1,\omega_2}\!\!\!\!\!\bra{vac}_{\omega}]\hat{E}^{(n)}\qty[\Motimes_{\omega\neq\omega_1,\omega_2}\!\!\!\!\!\ket{vac}_{\omega}]
\end{equation}
which acts in the subspace of modes $\omega_1$ and $\omega_2$. The trace over these modes is then
\begin{equation}\label{eq:dm_pk1}
    p^{(n)}=\frac{1}{M(M-1)} \sum_{\omega_2>\omega_1}{\Tr_{\omega_1,\omega_2}\qty(\hat{E}_{\omega_1,\omega_2}^{(n, \omega_1,\omega_2)}\qty[\hat\rho_a^{(\omega_1)}\hat\rho_b^{(\omega_2)}+\hat\rho_b^{(\omega_1)}\hat\rho_a^{(\omega_2)}])}.
\end{equation}

For any pair of modes $\omega_1$ and $\omega_2$, the corresponding two-mode state is the same and equals Eq.~(13) in the main text:
\begin{equation}\label{eq:rho2}
    \hat\rho_2 = \frac{1}{2} \qty( \hat\rho_a \otimes \hat\rho_b + \hat\rho_b \otimes \hat\rho_a),
\end{equation}
Thus, we can sum over the two-mode operators $\hat{E}_{\omega_1,\omega_2}^{(n)}$ as follows:

\begin{equation}
    \hat{E}_{2}^{(n)}=\frac{2}{M(M-1)} \sum_{\omega_2>\omega_1}{E_{\omega_1,\omega_2}^{(n)}},
\end{equation}
and obtain $p^{(n)}=\text{Tr}(\hat{E}_{2}^{(n)}\hat\rho_2)$.

Using the definition \eqref{eq:povm_w1_w2}, one can verify that $\hat{E}_{2}^{(n)} \geq 0$ for all $n$, and that $\sum_n{\hat{E}_{2}^{(n)}}=\hat I$. Thus, the operators $\{\hat{E}_{2}^{(n)}\}$ also form valid POVM effects.

Moreover, if we choose the effects $\hat{E}^{(n)}$ such that they maximize the Fisher information over the parameters encoded in $\rho_a$ and $\rho_b$, the operators $\hat{E}_{2}^{(n)}$ will also maximize it. Therefore, the quantum Fisher information for the states $\hat\rho_M$ and $\hat\rho_2$ is the same.

\subsection{Quantum Fisher information matrix calculation}\label{app:qfi}


Consider a quantum state of light $\hat \rho(\boldsymbol{\theta})$, depending on the unknown parameters $\boldsymbol{\theta} = \{d, \gamma, x_c, \mu \}$, with a spectral decomposition $\hat \rho = \sum_{j}{\lambda_j(\boldsymbol{\theta}) \ketbra{\psi_j(\boldsymbol{\theta})}}$. The quantum Fisher information matrix (qFIM) for this state can be calculated as \cite{Tsang_Nair_Lu_2015,Bisketzi_Branford_Datta_2019,Liu2020}:
\begin{equation}\label{eq:qfi}
    Q_{\alpha\beta} = 2 \sum_{k,l, \lambda_k + \lambda_l \neq 0} \frac{\mel{\psi_k}{\partial_\alpha\hat\rho}{\psi_l} \mel{\psi_l}{\partial_\beta\hat\rho}{\psi_k}}{\lambda_k + \lambda_l} \bigg|_{\boldsymbol\theta = \boldsymbol{\theta^\ast}},
\end{equation}
where $\boldsymbol{\theta^\ast}$ corresponds to the true parameter values.

The computation of the quantum Fisher information matrix for the state $\hat\rho_2$ (Eq.~13) is complicated due to the infinite number of spatial modes, each containing 0 or 1 photons. However, this task can be reduced by finding a minimal orthogonal basis that supports $\hat\rho_2$ and its derivatives. Following the main text, we define the single-photon state emitted from the source $s$ as
\begin{equation}
    \ket{\Psi_s} = \int{{\rm d}q \Psi_s(q) a^\dagger(q)}\ket{vac}, \qquad
    \Psi_s(q) = \sqrt{\frac{\sigma}{\sqrt{\pi}}} {\rm e}^{-\frac{\sigma^2q^2}{2}} {\rm e}^{-{\rm i}qx_s}.
\end{equation}

The partial derivative of this state with respect to some parameter $\theta$ is related to the $\HG{1}$ mode:
\begin{equation}
    \partial_\theta{\ket{\Psi_s}} = \frac{\partial_\theta{x_s}}{\sqrt{2}\sigma}\ket{\bar{\Psi}_s},
\end{equation}
where
\begin{equation}
    \ket{\bar{\Psi}_s} = \int{{\rm d}q \bar{\Psi}_s(q) a^\dagger(q)}\ket{vac}, \qquad
    \bar{\Psi}_s(q) = \sqrt{\frac{\sigma}{\sqrt\pi}}{\rm e}^{-\frac{\sigma^2q^2}{2}}{\rm e}^{-{\rm i}qx_s}\qty(-i\sqrt{2}\sigma q).
\end{equation}
Thus, the derivative of the single-source state $\hat\rho_s=(1-\mu_s)\ketbra{vac} + \mu_s\ketbra{\Psi_s}$ with respect to the parameter $\theta$ is
\begin{equation}
    \partial_\theta{\hat\rho_s}=\partial_\theta{\mu_s}\qty(\ketbra{\Psi_s}-\ketbra{vac}) + \frac{\mu_s\partial_\theta{x_s}}{\sqrt{2}\sigma}\qty(\ketbra{\Psi_s}{\bar{\Psi}_s}+\ketbra{\bar{\Psi}_s}{\Psi_s})
\end{equation}

The derivative of the two-mode state (Eq.~13) is
\begin{equation}
    \partial_\theta{\hat{\rho}_2}=\frac{1}{2} \qty( (\partial_\theta{\hat\rho_a}) \otimes \hat\rho_b + \hat\rho_a \otimes (\partial_\theta{\hat\rho_b}) + (\partial_\theta{\hat\rho_b}) \otimes \hat\rho_a + \hat\rho_b \otimes (\partial_\theta{\hat\rho_a})).
\end{equation}

For $\hat{\rho}_2$ and $\partial_\theta{\hat{\rho}_2}$, each mode is supported by the set of five states:
\begin{equation}\label{eq:state_set}
    \{\ket{vac}, \ket{\Psi_a}, \ket{\Psi_b}, \ket{\bar\Psi_a}, \ket{\bar\Psi_b}\}.
\end{equation}

Following \cite{Tsang_Nair_Lu_2015} (Appendix~C), we introduce an orthonormal basis $\{\ket{e_k},k=0,\dots,4\}$ such that:
\begin{equation}
\begin{gathered}
    \ket{vac} = \ket{e_0}, \quad \ket{\Psi_a} = a_+\ket{e_1} + a_-\ket{e_2}, \quad \ket{\Psi_b} = a_+\ket{e_1} - a_-\ket{e_2}, \\
    \ket{\bar{\Psi}_a}=-b_+\ket{e_1} + b_-\ket{e_2} + c_-\ket{e_3} + c_+\ket{e_4}, \\
    \ket{\bar{\Psi}_b}=+b_+\ket{e_1} + b_-\ket{e_2} + c_-\ket{e_3} - c_+\ket{e_4},
\end{gathered}
\end{equation}
where
\begin{equation}
    a_{\pm} = \sqrt{\frac{1 \pm V}{2}}, \quad b_{\pm} = \frac{d}{\sqrt{8}\sigma a_{\pm}}V, \quad c_{\pm}=\sqrt{a_{\mp}^2 \pm \frac{b_{\pm}^2}{V}},
\end{equation}
and $V = {\rm e}^{-\frac{d^2}{4\sigma^2}}$. Using this decomposition, we define $\hat{\rho}_2$ and $\partial_\theta{\hat{\rho}_2}$ in the matrix representation and then compute the quantum Fisher information matrix numerically.

\newpage
\section{Comparison of error bounds}
\subsection{Analytical approximations for $d\rightarrow 0$}
\label{ss:CRBs.analytic}

In Table~\ref{tab:ESDs}, we present semi-analytical equations for the estimation variances of the parameters at the limit $d \rightarrow 0$, assuming small values of $x_c$ and $\mu$. As can be seen, for all parameters except $d$, both ESS and BLESS protocols yield estimation variances similar to qCRB-2 (up to a constant factor). However, only the BLESS protocol reaches qCRB-2 for distance estimation, while the ESS protocol gives a distance variance that is larger by a factor of $\gamma^{-2}\mu^{-1}$.

\begin{table}[htb]
    \centering
    \renewcommand{\arraystretch}{1.8} 
    \begin{tabular}{|c|c|c|c|c|}
    \hline
&$N \Delta_d^2$ &$ N \Delta_\gamma^2$ &$ N \Delta_{x_c}^2$ &$ N \Delta_\mu^2$ \\ \hline 
 ESS & $\frac{9.6 \sigma^2}{\gamma^2 g \mu^2}$ &  $\frac{3 g}{\gamma^2  \mu^2}$ &  $\frac{3.5\sigma^2}{\mu}$& $1.3 \mu$\\ \hline
  BLESS & $ \frac{5.3 \sigma^2 }{g \mu}$ &  $\frac{5.6 g}{\gamma^2  \mu^2}$ & $\frac{7.0\sigma^2}{\mu}$ & $2.6 \mu$\\ \hline
    qCRB-2 & $ \frac{\sigma^2 }{g \mu}$ &  $\frac{ 2 g}{\gamma^2  \mu^2}$ & $\frac{ \sigma^2}{2 \mu}$ & $ \mu$\\ \hline
    \end{tabular}
    \caption{Estimation variances of all the model parameters at the limit $d\rightarrow 0$, assuming $x_c\ll\sigma$ and $\mu\ll 1$. }
    \label{tab:ESDs}
\end{table}

\subsection{Exposure time needed for 10-fold resolution enhancement}
\label{ss:Time}

Define the resolution enhancement factor with respect to the Rayleigh limit as $\alpha \equiv \sigma/d$. Figures~\ref{fig:ESDs}~(a,g) show the normalized distance estimation error $\Delta_d\sqrt{N}/\sigma \equiv \kappa$ versus $\alpha^{-1}$. The minimal exposure time needed to achieve a given enhancement $\alpha$ (i.e., obtain $\Delta_d \leq d$) is $T(\alpha) = \tau N = \tau \kappa^2 \alpha^2$, where $\tau$ is the single measurement time, which is equal to the luminescence lifetime.

Notably, even under conditions that break Rayleigh's curse ($\kappa = \text{const}$), the exposure time $T$ scales as $T \propto \alpha^2$, while for non-optimal imaging protocols, it scales as $T \propto \alpha^4$ or even $T \propto \alpha^6$.

Exposing times needed to achieve a 10-fold resolution enhancement $T(\alpha=10)$ for all measurement protocols are presented in Table~\ref{tab:Times}. All values, except for BLESS with imperfections, were calculated from Fig.~3~(a). The last value was calculated from Fig.~3~(g).

\begin{table}[htb]
    \centering
    \renewcommand{\arraystretch}{1.8} 
    \begin{tabular}{|c|c|}
    \hline
Measurement protocol & Exposing time $T(\alpha=10)$\\
\hline
Quantum limit (qCRB-2)& 1~ms\\ \hline
Direct Imaging (DI) and SPADE & 500~s\\ \hline
ESS protocol & 5~s\\ \hline
BLESS protocol & 5~ms \\ \hline
BLESS with imperfections & 50~ms \\ \hline
    \end{tabular}
    \caption{Exposure time needed to obtain a 10-fold resolution enhancement for various measurement protocols.}
    \label{tab:Times}
\end{table}

\bibliographystyle{plain}
\bibliography{article}